\documentclass[a4paper,11pt]{article}

\pdfoutput=1 

\usepackage{jheppub} 

\usepackage{slashed}
\usepackage{subcaption}
\usepackage{xcolor}
\usepackage{cleveref}
\usepackage[T1]{fontenc} 
\usepackage[utf8]{inputenc}
\usepackage{mathtools}

\DeclarePairedDelimiter\floor{\lfloor}{\rfloor}

\newcommand{\cM}{\mathcal{M}}

\newcommand{\cF}{\mathcal{F}}

\newcommand{\cO}{\mathcal{O}}

\newcommand{\wa}{ w \cdot \mathfrak{a} }

\newcommand{\bv}{\boldsymbol{v}}
\newcommand{\bbv}{\bar{\boldsymbol{v}}}

\makeatletter
\newsavebox{\@brx}
\newcommand{\llangle}[1][]{\savebox{\@brx}{\(\m@th{#1\langle}\)}%
  \mathopen{\copy\@brx\kern-0.5\wd\@brx\usebox{\@brx}}}
\newcommand{\rrangle}[1][]{\savebox{\@brx}{\(\m@th{#1\rangle}\)}%
  \mathclose{\copy\@brx\kern-0.5\wd\@brx\usebox{\@brx}}}
\makeatother

\subheader{\normalsize
	\vspace*{-3.7em}
	\begin{flushright}
		UUITP-02/23
	\end{flushright}
}

\title{\boldmath Recursion in the classical limit and the neutron-star Compton amplitude}

\author[a,b]{Kays Haddad}
\affiliation[a]{Department of Physics and Astronomy, Uppsala University, \\
Box 516, 75120 Uppsala, Sweden}
\affiliation[b]{Nordita, Stockholm University and KTH Royal Institute of Technology, \\
Hannes Alfv\'{e}ns v\"{a}g 12, 10691 Stockholm, Sweden}
\emailAdd{kays.haddad@physics.uu.se}

\abstract{We study the compatibility of recursive techniques with the classical limit of scattering amplitudes through the construction of the classical Compton amplitude for general spinning compact objects.
This is done using BCFW recursion on three-point amplitudes expressed in terms of the classical spin vector and tensor, and expanded to next-to-leading-order in $\hbar$ by using the heavy on-shell spinors.
Matching to the result of classical computations, we find that lower-point quantum contributions are, in general, required for the recursive construction of classical, spinning, higher-point amplitudes with massive propagators.
We are thus led to conclude that BCFW recursion and the classical limit do not commute.
In possession of the classical Compton amplitude, we remove non-localities to all orders in spin for opposite graviton helicities, and to fifth order in the same-helicity case.
Finally, all possible on-shell contact terms potentially relevant to black-hole scattering at the second post-Minkowskian order are enumerated and written explicitly.
}

\allowdisplaybreaks

\begin{document} 
\maketitle
\flushbottom

\section{Introduction}

Since the first observations of gravitational waves from compact binary coalescence \cite{LIGOScientific:2016aoc,LIGOScientific:2017vwq,LIGOScientific:2018mvr}, massive effort from the scattering-amplitudes community has been dedicated to understanding the link between quantum scattering amplitudes and classical gravitational phenomena.
Several of these works have improved the (post-Minkowskian, PM) precision to which we understand compact binaries \cite{Cheung:2018wkq,Bern:2019nnu,Bern:2019crd,Cheung:2020gyp,Kalin:2020fhe,Bern:2021dqo,Dlapa:2021npj,Bern:2021yeh,Dlapa:2021vgp,Jakobsen:2022fcj}, including in the presence of additional effects such as spin  \cite{Guevara:2018wpp,Chung:2018kqs,Maybee:2019jus,Guevara:2019fsj,Damgaard:2019lfh,Aoude:2020onz,Bern:2020buy,Liu:2021zxr,Kosmopoulos:2021zoq,Jakobsen:2021lvp,Jakobsen:2021zvh,Chen:2021kxt,Aoude:2022trd,Bern:2022kto,Aoude:2022thd,FebresCordero:2022jts}, radiation \cite{AccettulliHuber:2020dal,Herrmann:2021lqe,DiVecchia:2021bdo,Herrmann:2021tct,Bjerrum-Bohr:2021din,Alessio:2022kwv,Jakobsen:2022psy,Jakobsen:2022zsx}, tidal effects \cite{Cheung:2020sdj,Haddad:2020que,Kalin:2020lmz,Bern:2020uwk,Cheung:2020gbf,Aoude:2020ygw,Heissenberg:2022tsn}, and various combinations thereof.
Equally as foundational has been the derivation of connections between scattering amplitudes and classical observables \cite{Cheung:2018wkq,Kosower:2018adc,Cristofoli:2019neg,Maybee:2019jus,Kalin:2019rwq,Bjerrum-Bohr:2019kec,Kalin:2019inp,Cristofoli:2021vyo,Bautista:2021wfy,Aoude:2021oqj,Cho:2021arx,Adamo:2022rmp,Bautista:2022wjf}.

Many developments emerging from this mobilization have centered on or been motivated by the need for the efficient extraction of the classically-relevant portion of a scattering amplitude.
Along these lines, there have been works involving convenient $\hbar$-counting schemes \cite{Kosower:2018adc,Maybee:2019jus}, Lagrangian-level as well as spinor-helicity heavy/classical limits \cite{Damgaard:2019lfh,Aoude:2020onz,Brandhuber:2021kpo,Brandhuber:2021eyq,Bjerrum-Bohr:2023jau}, effective field theories (EFTs) with classical spin degrees of freedom (DoFs) \cite{Bern:2020buy,Bern:2022kto}, and a worldline EFT obtained by integrating out quantum DoFs \cite{Mogull:2020sak,Jakobsen:2021zvh}.
The efficient computation of classically-relevant amplitudes is a central inspiration for the present work.
Specifically, we attempt to qualify the compatibility of BCFW recursion \cite{Britto:2004ap,Britto:2005fq} with the classical limit, seeking instances of potential simplifications to the recursive construction of higher-multiplicity amplitudes in this limit.
Our conduit for this study is the classical Compton amplitude.

While much focus has recently been dedicated to the appropriate Compton amplitude for describing Kerr black holes \cite{Arkani-Hamed:2017jhn,Guevara:2018wpp,Chung:2018kqs,Falkowski:2020aso,Bautista:2021wfy,Chiodaroli:2021eug,Aoude:2022thd,Bern:2022kto,Cangemi:2022bew,Bautista:2022wjf}, in this paper we focus on deriving this amplitude for a general spinning compact object.
Though conceptually simpler -- one must "only" enumerate all possible Wilson coefficients and leave their values unspecified -- this task is computationally more subtle.
Certain serendipitous cancellations occur in the recursive construction of the black-hole Compton amplitude,\footnote{In the rest of the paper, we use "black-hole Compton amplitude" and "black-hole limit" to refer to the amplitude with the black-hole values for the linear-in-curvature, spin-induced multipole coefficients, $C_{S^{j}}=1$. We are not concerned with the values of $R^{2}$ Wilson coefficients that describe black holes.} allowing one to build this amplitude using only leading-in-$\hbar$ information throughout the computation.
As we will see below, this fortune does not extend to the case of general objects, where subleading-in-$\hbar$ information is needed in intermediate steps to match to known classical results \cite{Saketh:2022wap}.

Apart from phenomenological applications, there are two reasons why an on-shell study of the Compton amplitude for general compact objects is itself timely.
The first is the discrepancy between the number of free parameters in the spinning effective field theory of refs.~\cite{Bern:2020buy,Bern:2022kto} and worldline theories of spinning objects \cite{Levi:2015msa,Siemonsen:2019dsu,Jakobsen:2021zvh} at linear order in the curvature (but contributing first to the Compton amplitude).
An on-shell perspective provides a different take on the total possible number of free parameters, in a setting where relations between different structures may be easier to identify than in the off-shell context.
Recently, ref.~\cite{Kim:2023drc} argued that the actions of refs.~\cite{Bern:2020buy,Bern:2022kto} possess too many degrees of freedom due to incompletely removing unphysical massive modes.
Our on-shell analysis corroborates the conclusion of ref.~\cite{Kim:2023drc}, in that we find no freedom to introduce coefficients to the three-point amplitude other than those mapping directly to the parameters of ref.~\cite{Levi:2015msa}, including at next-to-leading order in $\hbar$.

The second reason is the burgeoning interest in and necessity for an amplitudes description of Kerr black holes.
Extrapolating the properties of Kerr-black-hole scattering at low spin orders, it was proposed in refs.~\cite{Aoude:2022trd,Bern:2022kto,Aoude:2022thd} that higher spin orders in Kerr-black-hole scattering would exhibit a favorable high-energy limit and a so-called spin-shift symmetry.
However, these conjectures are in tension with recent comparisons to solutions of the Teukolsky equation \cite{Bautista:2022wjf}, indicating a need for a better understanding of the amplitudes pertinent to Kerr black holes.\footnote{Refs.~\cite{Chiodaroli:2021eug,Cangemi:2022bew} have proposed that Kerr-black-hole amplitudes exhibit a massive higher-spin gauge symmetry. This gauge symmetry uniquely selects the Kerr three-point amplitude, but the dependence of their Compton amplitudes on the spin quantum number of the massive particle necessitates a more careful infinite-spin limit before comparisons to known Kerr Compton amplitudes can be made. We thank Lucile Cangemi, Henrik Johansson, and Paolo Pichini for clarifications about this.}
Computation of the most-general Compton amplitude works towards this end by allowing for comparisons to be made to the computation relevant for the Kerr Compton amplitude.

The Compton amplitude for general spin-induced multipoles has been computed recursively in ref.~\cite{Chen:2021kxt}.
We do not fully agree with their results, and address this in more detail below.
Put succinctly, the cause of the disagreement -- as well as a conclusion of our analysis -- is that the classical limit and BCFW recursion generally \textit{do not commute}.
Our analysis here postpones the full classical limit until the end of the computation, and produces the opposite-helicity Compton amplitude for general spin-induced multipoles to all spin orders and without any unphysical poles. 
The result matches the classical computation of ref.~\cite{Saketh:2022wap} where there is overlap.
In the same-helicity case we must cure a new type of non-locality appearing already at quadratic order in spin, which we do up to fifth order in spin and again reproduce the results of ref.~\cite{Saketh:2022wap} where there is overlap.
Beyond fifth order in spin, a non-local form of the same-helicity amplitude is presented for all spin orders.
Consideration of the massless limit in the same-helicity case will hint at a potential extension of the notion of minimal coupling of ref.~\cite{Arkani-Hamed:2017jhn} past three points.

Apart from the linear-in-curvature spin-induced multipole coefficients, the identity of the compact object in the Compton amplitude is also dictated by the values for coefficients of contact terms, corresponding to $R^{2}$ operators in an action.
Due to the presence of more non-vanishing invariants than at three points, there are an infinite number of such coefficients at each spin order.
Once we have constructed the factorizable portion of the Compton amplitude, we content ourselves with the enumeration of a certain finite subset of all possible contact deformations of the Compton amplitude -- specifically, the set of terms which can potentially contribute to Kerr-black-hole scattering at 2PM.

The remainder of this paper is organized as follows.
In \Cref{sec:ThreePoints} we expand the most-general three-point amplitude up to subleading order in $\hbar$ while expressing it in terms of classically relevant quantities: the spin vector, the spin tensor, and the linear-in-curvature spin-induced multipole coefficients.
The three-point amplitude thus expanded is sufficient for the computation of the factorizable part of the Compton amplitude, which we carry out for all helicity configurations in \Cref{sec:Compton}.
Contact deformations of the Compton amplitudes of the variety mentioned above are counted and written explicitly in \Cref{sec:Contact}.
The analysis in this section includes both conservative and dissipative contact terms.
A discussion of our results and their implications concludes the paper in \Cref{sec:Summary}.

\section{The most-general three-point amplitude}\label{sec:ThreePoints}

As our analysis hinges on the application of BCFW recursion to construct the Compton amplitude, our starting point is the expression of the three-point amplitude in terms of classically-relevant quantities.
The most-general spin-$s$ three-point amplitude in terms of the massive on-shell spinors of ref.~\cite{Arkani-Hamed:2017jhn} is \cite{Conde:2016izb,Conde:2016vxs,Arkani-Hamed:2017jhn}
\begin{align}
    \left(\cM^{s,+}_{3}\right)^{IJ}&=i\frac{x^{2}}{m^{2s}}\sum_{k=0}^{2s}g_{k}\langle2^{I}1^{J}\rangle^{\odot(2s-k)}\odot\left(\frac{\langle2^{I}|q|1^{J}]}{2m}\right)^{\odot k},\label{eq:ThreePointPos} \\
    \left(\cM^{s,-}_{3}\right)^{IJ}&=i(-1)^{2s}\frac{x^{-2}}{m^{2s}}\sum_{k=0}^{2s}\tilde{g}_{k}[2^{I}1^{J}]^{\odot(2s-k)}\odot\left(\frac{[2^{I}|q|1^{J}\rangle}{2m}\right)^{\odot k},\label{eq:ThreePointNeg}
\end{align}
for an emitted graviton of any helicity.
The helicity weights of the gravitons are encoded in the factors
\begin{align}
    x\equiv\frac{[q|p_{1}|\xi\rangle}{m\langle q\xi\rangle},\quad x^{-1}\equiv\frac{\langle q|p_{1}|\xi]}{m[q\xi]},
\end{align}
for an arbitrary reference vector $\xi^{\mu}$.
Amplitudes describing the scattering of massive particles with spins $s_{i}$ are symmetric functions in the $2s_{i}$ little group indices of each massive spinning particle \cite{Wigner:1939cj,Bargmann:1948ck,weinberg_1995ch2}.
The $I=\{I_{1},\dots,I_{2s}\}$ and $J=\{J_{1},\dots,J_{2s}\}$ represent these sets of $2s$ indices for the outgoing and incoming massive legs, respectively.
We have used the $\odot$ notation first introduced in ref.~\cite{Guevara:2018wpp} to represent the symmetrization of the tensor product over the little group indices.\footnote{For example, ${x^{I}}_{J}\odot {y^{I}}_{J}={x^{\{I_{1}}}_{\{J_{2}}{y^{I_{2}\}}}_{J_{2}\}}$, where curly brackets denote normalized symmetrization.}
Parameters of classical relevance will shortly be introduced with which we will identify the coefficients $g_{k}$ and $\tilde{g}_{k}$.

To express these in terms of the classical spin vector up to subleading order in $\hbar$, we will convert the on-shell spinors to heavy on-shell spinors \cite{Aoude:2020onz,Aoude:2022trd}.
For a momentum $p^{\mu}=mv^{\mu}+k^{\mu}$,
\begin{align}
    |p^{I}\rangle=\frac{m}{\sqrt{m_{k}}}\left(|v^{I}\rangle+\frac{\slashed{k}}{2m}|v^{I}]\right),&\quad |p^{I}]=\frac{m}{\sqrt{m_{k}}}\left(|v^{I}]+\frac{\slashed{k}}{2m}|v^{I}\rangle\right), \\
    \langle p^{I}|=\frac{m}{\sqrt{m_{k}}}\left(\langle v^{I}|-[v^{I}|\frac{\slashed{k}}{2m}\right),&\quad [p^{I}|=\frac{m}{\sqrt{m_{k}}}\left([v^{I}|-\langle v^{I}|\frac{\slashed{k}}{2m}\right),
\end{align}
where $m_{k}\equiv\left(1-\frac{k^{2}}{4m^{2}}\right)m$.
Note that the residual momentum $k^{\mu}\sim\cO(\hbar)$ \cite{Damgaard:2019lfh}.
Writing $p_{1}^{\mu}=mv^{\mu}+k_{1}^{\mu}$ and $p_{2}^{\mu}=mv^{\mu}+k_{2}^{\mu}=mv^{\mu}+k_{1}^{\mu}-q^{\mu}$, the spinor brackets are
\begin{align}
    \langle2^{I}1^{J}\rangle&=m\left(\langle v^{I}v^{J}\rangle+\langle v^{I}|q\cdot a_{1/2}|v^{J}\rangle-\frac{i}{2m^{2}}k_{1\mu}q_{\nu}\langle v^{I}|S^{\mu\nu}_{1/2}|v^{J}\rangle\right)+\cO(\hbar^{2}),\notag \\
    [2^{I}1^{J}]&=m\left([v^{I}v^{J}]-[v^{I}|q\cdot a_{1/2}|v^{J}]-\frac{i}{2m^{2}}k_{1\mu}q_{\nu}[v^{I}|S^{\mu\nu}_{1/2}|v^{J}]\right)+\cO(\hbar^{2}),\label{eq:HeavySpinorConversions} \\
    \frac{\langle2^{I}|q|1^{J}]}{2m}&=m\langle v^{I}|q\cdot a_{1/2}|v^{I}\rangle+\cO(\hbar^{2}),\quad \frac{[2^{I}|q|1^{J}\rangle}{2m}=-m[v^{I}|q\cdot a_{1/2}|v^{I}]+\cO(\hbar^{2}),\notag
\end{align}
where $S^{\mu\nu}_{1/2}\equiv\frac{i}{4}[\gamma^{\mu},\gamma^{\nu}]$ is the Lorentz generator in the spin-1/2 representation, and we have indicated that the ring radius $a^{\mu}$ is accordingly in the spin-1/2 representation.
The ring radius is related to the spin vector through $S^{\mu}=a^{\mu}/m$; see \Cref{app:SpinProperties} for our conventions pertaining to the ring radius, as well as some of its germane properties.

With \cref{eq:HeavySpinorConversions} in hand, we can express the three-point amplitudes in terms of the classical spin.
Focusing on the positive-helicity amplitude and expanding up to next-to-leading-order in $\hbar$,
\begin{align}
    \left(\cM^{s,+}_{3}\right)^{IJ}&=ix^{2}\sum_{j=0}^{2s}\langle v^{I}v^{J}\rangle^{\odot(2s-j)}\odot\langle v^{I}|q\cdot a|v^{J}\rangle^{\odot(j-1)}\odot\left(\langle v^{I}|q\cdot a|v^{J}\rangle\sum_{k=0}^{j}\frac{g_{k}(2s-k)!}{(2s)!(j-k)!}\right.\notag \\
    &\quad\left.-\frac{i}{2m^{2}}k_{1\mu}q_{\nu}\langle v^{I}|S^{\mu\nu}|v^{J}\rangle\sum_{k=0}^{j}\frac{g_{k}(2s-k)!}{(2s)!(j-k-1)!}\right)+\cO(\hbar^{2}),
\end{align}
where we interpret $1/(-1)!=0$.
Note that we have converted the products of the ring radius and the spin tensor to the spin-$s$ representation, which we denote with no subscript on these quantities.

Now, the coefficients $g_{i}$ are not classically relevant,\footnote{As such, we do not concern ourselves with the fact that they must depend on the total spin quantum number to preserve spin universality.} but they can be related to linear combinations of the spin-induced multipole coefficients $C_{ES^{2k}}$ and $C_{BS^{2k+1}}$ of ref.~\cite{Levi:2015msa}.
We adopt the notation of ref.~\cite{Chung:2018kqs} with respect to these coefficients, writing $C_{S^{2k}}\equiv C_{ES^{2k}}$ and $C_{S^{2k+1}}\equiv C_{BS^{2k+1}}$.
Matching to the three-point amplitude that can be derived from the worldline action there, it's straightforward to show that\footnote{See ref.~\cite{Chung:2018kqs} for details of the extraction of an on-shell three-point amplitude from the worldline action of ref.~\cite{Levi:2015msa}. Ref.~\cite{Chen:2021kxt} finds similar expressions relating the amplitude and spinning-worldline coefficients.}
\begin{align}
    \sum_{k=0}^{j}g_{k}\frac{(2s-k)!}{(2s)!(j-k)!}= \frac{C_{S^{j}}}{j!}\quad\Rightarrow\quad g_{k}=\frac{(2s)!}{(2s-k)!}\sum_{n=0}^{k}(-1)^{n+k}\frac{C_{S^{n}}}{n!(k-n)!}.
\end{align}
Consequently,
\begin{align}
    &\left(\cM^{s,+}_{3}\right)^{IJ}
    =ix^{2}\left(\sum_{j=0}^{2s}\frac{C_{S^{j}}}{j!}\langle v^{I}v^{J}\rangle^{\odot2s-j}\odot\langle v^{I}|q\cdot a|v^{J}\rangle^{\odot j}\right. \\
    &\qquad\left.-\frac{i}{2m^{2}}k_{1\mu}q_{\nu}\langle v^{I}|S^{\mu\nu}|v^{J}\rangle\odot\sum_{j=1}^{2s}\frac{C_{S^{j-1}}}{(j-1)!}\langle v^{I}v^{J}\rangle^{\odot2s-j}\odot\langle v^{I}|q\cdot a|v^{J}\rangle^{\odot(j-1)}\right)+\cO(\hbar^{2}).\notag
\end{align}
The two coefficients $C_{S^{0}}$ and $C_{S^{1}}$ are equal to $1$ for any object, while the coefficients $C_{S^{j\geq2}}$ are all equal to $1$ for black holes only \cite{Levi:2015msa}.

This form of the amplitude with open little group indices is important to correctly account for polarization sums over massive internal states when computing the Compton amplitude recursively.
Nevertheless, the amplitude can be compactified by using the bold notation as formulated in ref.~\cite{Chiodaroli:2021eug} and employed in ref.~\cite{Aoude:2022trd}:
\begin{align*}
    |\bv\rangle\equiv|v^{I}\rangle z_{p,I},\quad |\bv]\equiv|v^{I}] z_{p,I}, \\
    \langle\bbv|\equiv\bar{z}_{p,I}\langle v^{I}|,\quad [\bbv|\equiv \bar{z}_{p,I}[v^{I}|,
\end{align*}
where $z_{p,I}$ is a complex auxiliary variable and $\bar{z}_{p,I}$ its complex conjugate.
Contracting the amplitude with $2s$ factors of this auxiliary variable for each massive leg,
\begin{align}\label{eq:ThreePointPositive}
    \cM^{s,+}_{3}&=ix^{2}\langle\bbv\bv\rangle^{2s}\left(\sum_{j=0}^{2s}\frac{C_{S^{j}}}{j!}\left(q\cdot\mathfrak{a}\right)^{j}-\frac{i}{2m^{2}}k_{1\mu}q_{\nu}\mathfrak{s}^{\mu\nu}\sum_{j=0}^{2s-1}\frac{C_{S^{j}}}{j!}(q\cdot\mathfrak{a})^{j}\right)+\cO(\hbar^{2}).
\end{align}
We have identified (products of) the \textit{classical} ring radius and spin tensor --  $\mathfrak{a}^{\mu}$ and $\mathfrak{s}^{\mu\nu}$ respectively -- through \cite{Bern:2020buy,Cangemi:2022abk}\footnote{The symmetrization of the spin in the expectation value was not important in ref.~\cite{Aoude:2022trd} because the product of spin vectors there was contracted with a totally-symmetric tensor when identified in the amplitude.}
\begin{align}
    \mathfrak{a}^{\mu_{1}}\dots\mathfrak{a}^{\mu_{n}}&\equiv\frac{\langle\bbv|^{2s}\{a^{\mu_{1}},\ldots,a^{\mu_{n}}\}|\bv\rangle^{2s}}{\langle\bbv\bv\rangle^{2s}}=\frac{[\bbv|^{2s}\{a^{\mu_{1}},\ldots,a^{\mu_{n}}\}|\bv]^{2s}}{[\bbv\bv]^{2s}},\\
    \mathfrak{s}^{\alpha\beta}\mathfrak{a}^{\mu_{1}}\dots\mathfrak{a}^{\mu_{n}}&\equiv\frac{\langle\bbv|^{2s}\{S^{\alpha\beta},a^{\mu_{1}},\ldots,a^{\mu_{n}}\}|\bv\rangle^{2s}}{\langle\bbv\bv\rangle^{2s}}=\frac{[\bbv|^{2s}\{S^{\alpha\beta},a^{\mu_{1}},\ldots,a^{\mu_{n}}\}|\bv]^{2s}}{[\bbv\bv]^{2s}}.
\end{align}
Finally, once in possession of the positive-helicity amplitude, its negative-helicity counterpart is easily obtained by swapping angle and square brackets and changing the sign on $q\cdot\mathfrak{a}$, as can be seen by inspection of \cref{eq:HeavySpinorConversions}:
\begin{align}\label{eq:ThreePointNegative}
    \cM^{s,-}_{3}&=ix^{-2}\langle\bbv\bv\rangle^{2s}\left(\sum_{j=0}^{2s}\frac{C_{S^{j}}}{j!}\left(-q\cdot\mathfrak{a}\right)^{j}-\frac{i}{2m^{2}}k_{1\mu}q_{\nu}\mathfrak{s}^{\mu\nu}\sum_{j=0}^{2s-1}\frac{C_{S^{j}}}{j!}(-q\cdot\mathfrak{a})^{j}\right)+\cO(\hbar^{2}).
\end{align}
Note that we switched back from square to angle brackets to absorb the overall $(-1)^{2s}$ in \cref{eq:ThreePointNeg}.
In the infinite-spin limit, one should drop the overall factors of $\langle\bbv\bv\rangle^{2s}$ and take $s\rightarrow\infty$ in the upper bounds of the sums in \cref{eq:ThreePointPositive,eq:ThreePointNegative}.
However, we must postpone this procedure until after the Compton amplitudes have been constructed for arbitrary, but finite, $s$: the presence of spinors is necessary to correctly perform massive polarization sums in intermediate steps.

The second terms in the brackets of both \cref{eq:ThreePointPositive,eq:ThreePointNegative} are subleading in $\hbar$.
While they are needed to correctly construct the classical Compton amplitude, they are irrelevant to classical physics at three points.\footnote{One could also set $k_{1}^{\mu}=0$ by reparametrizing the heavy momentum. This not only eliminates the second terms in the brackets of \cref{eq:ThreePointPositive,eq:ThreePointNegative}, but sets \textit{all} subleading-in-$\hbar$ terms to zero.}
Indeed, dropping these terms and taking the black-hole limit $C_{S^{j}}=1$, we recover the spin exponential characteristic of Kerr black holes at three points \cite{Vines:2017hyw,Guevara:2018wpp}.
We thus see the potential to introduce new Wilson coefficients that would not affect the classical three-point amplitude but would enter in the classical Compton amplitude.
Specifically, a bottom-up construction of a three-point amplitude in powers of $\hbar$ would require that we give the second terms in the brackets of \cref{eq:ThreePointPositive,eq:ThreePointNegative} coefficients different from $C_{S^{j}}$.
However, knowledge of the underlying -- quantum -- theory indicates that there are in fact no additional parameters to the $C_{S^{j}}$ if we are to match to the classical three-point amplitude, as we have derived above.
One would reach the same conclusion from the bottom up if, in addition to enumerating all possible on-shell structures, one also imposes invariance of the amplitude under reparametrization of the heavy momentum \cite{DUGAN1992142,Luke:1992cs}.

We have derived the three-point amplitudes for the emission of an arbitrary-helicity graviton from a spin-$s$ massive particle up to subleading order in $\hbar$ and in terms of classically relevant quantities (the ring radius, spin tensor, and spin-induced multipole coefficients).
This is all the input we need to construct the classical Compton amplitude using recursive methods.

\section{BCFW construction of Compton scattering}\label{sec:Compton}

We turn now to the sewing of the three-point amplitudes derived in the previous section into Compton amplitudes.
Inspection of the factorization channels will validate our assertion that BCFW recursion and the classical limit do not commute, as we will see interference between quantum and superclassical terms that generally does not vanish.
We begin with the opposite-helicity case before proceeding to same-helicity scattering,\footnote{Note that the opposite- and same-helicity amplitudes are often referred to as helicity-preserving and helicity-reversing amplitudes respectively in the general relativity literature, e.g. refs.~\cite{Dolan:2008kf,Saketh:2022wap}. This nomenclature reflects momentum conventions where one graviton is incoming and one is outgoing.} and take both gravitons to be outgoing and the initial matter momentum to be incoming.

\subsection{Opposite-helicity Compton scattering}

\begin{figure}
\centering
\subfloat{
    \includegraphics[scale=0.3]{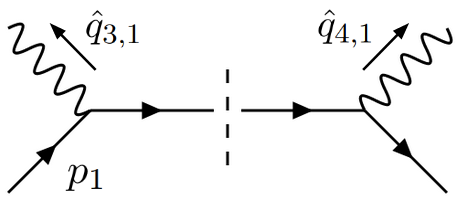}
    }
\hspace{1cm}
\subfloat{
    \includegraphics[scale=0.3]{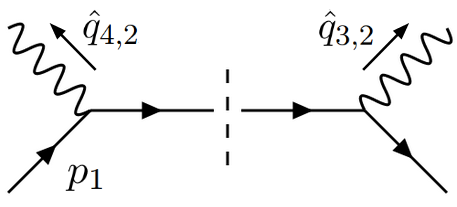}
    }
\caption{\label{fig:OppHelFactorizations}The two factorization channels composing the spin-$s$ opposite-helicity Compton amplitude under the $[3,4\rangle$ shift. We label the factorization channel on the left $\cF_{1}^{s,-+}$ and that on the right $\cF_{2}^{s,-+}$.}
\end{figure}

We label the negative-helicity graviton's momentum with $q_{3}^{\mu}$, while the positive-helicity graviton has momentum $q_{4}^{\mu}$.
We use the $[3,4\rangle$ shift to construct the amplitude:
\begin{align}
    |\hat{3}]=|3]+z|4],\quad |\hat{4}\rangle=|4\rangle-z|3\rangle.
\end{align}
Two factorization channels comprise the amplitude under this shift, which are shown in \cref{fig:OppHelFactorizations}.
On either cut, the factors of $x$ in \cref{eq:ThreePointPositive,eq:ThreePointNegative} can be written as
\begin{align}
    \hat{x}_{4}=\frac{y}{m\langle43\rangle},\quad \hat{x}_{3}^{-1}=\frac{y}{m[34]},
\end{align}
where $y\equiv2p_{1}\cdot w$ and $w^{\mu}\equiv[4|\bar{\sigma}^{\mu}|3\rangle/2$.
The inverse powers of the mass will cancel with the overall coupling, so we omit them in the following.
The shifted momenta on the cuts are
\begin{align}\label{eq:OppositeHelicityShiftedMomenta}
    \hat{q}_{3,1}^{\mu}&=\hat{q}_{3}^{\mu}-\frac{s_{34}}{2y}w^{\mu},\quad \hat{q}_{3,2}^{\mu}=\hat{q}_{3}^{\mu}+\frac{s_{34}}{2y}w^{\mu}, \\
    \hat{q}_{4,1}^{\mu}&=\hat{q}_{4}^{\mu}+\frac{s_{34}}{2y}w^{\mu},\quad \hat{q}_{4,2}^{\mu}=\hat{q}^{\mu}_{4}-\frac{s_{34}}{2y}w^{\mu},
\end{align}
where
\begin{align}\label{eq:UnshiftedHats}
    \hat{q}_{3}^{\mu}\equiv q_{3}^{\mu}-\frac{t_{14}-t_{13}}{2y}w^{\mu},\quad \hat{q}_{4}^{\mu}\equiv q_{4}^{\mu}+\frac{t_{14}-t_{13}}{2y}w^{\mu}.
\end{align}
Summing the two factorization channels yields the spin-$s$ amplitude:
\begin{align}
    \cM^{s,-+}_{4}&=\cF_{1}^{s,-+}+\cF_{2}^{s,-+}, \\
    \cF_{1}^{s,-+}&\equiv\frac{\cM_{R}(\hat{P}_{13,I}^{s},-\hat{q}_{4,1}^{+},-p_{2}^{s})\cM_{L}(p_{1}^{s},-\hat{q}_{3,1}^{-},-\hat{P}_{13}^{s,I})}{t_{13}},\notag \\
    \cF_{2}^{s,-+}&\equiv\frac{\cM_{R}(\hat{P}_{14,I}^{s},-\hat{q}_{3,2}^{-},-p_{2}^{s})\cM_{L}(p_{1}^{s},-\hat{q}_{4,2}^{+},-\hat{P}_{14}^{s,I})}{t_{14}}.\notag
\end{align}
The labels $R$ and $L$ indicate whether the amplitude is on the right- or left-hand side of the cut, which affects the sign of the polarization sum.
We have defined $\hat{P}_{13}=p_{1}^{\mu}-\hat{q}_{3,1}^{\mu}$, $\hat{P}_{14}=p_{1}^{\mu}-\hat{q}_{4,2}^{\mu}$, $t_{1i}=-2p_{1}\cdot q_{i}$, and $p_{2}^{\mu}$ is determined by momentum conservation.
Negative momentum labels in the amplitudes represent outgoing momenta.

Since the three-point amplitudes are $\cO(\hbar^{0})$ we see that both factorization channels are $\cO(\hbar^{-1})$.
From previous analyses \cite{Arkani-Hamed:2017jhn,Chung:2018kqs,Guevara:2018wpp,Johansson:2019dnu,Aoude:2020onz,Aoude:2022trd} it is known that the Compton amplitude scales as $\cO(\hbar^{0})$ in the classical limit, so $\cO(\hbar)$ terms from the three-point amplitudes are needed to capture all $\cO(\hbar^{0})$ contributions to the Compton amplitude, as advertised.
On the left-hand sides of the cuts we can take $k_{1}^{\mu}=0$ in \cref{eq:ThreePointPositive,eq:ThreePointNegative}, but on the right-hand sides we must take $k_{1}^{\mu}=-\hat{q}_{3,1}^{\mu}$ or $-\hat{q}_{4,2}^{\mu}$, depending on the cut.

We see in \cref{eq:OppositeHelicityShiftedMomenta} the second source of $\cO(\hbar\times 1/\hbar)$ effects.
Namely, the parts of the shifted momenta proportional to $w^{\mu}$ are $\cO(\hbar^{2})$ whereas the rest of the shifted momenta are $\cO(\hbar)$.
There is one more source of interference between quantum and super-classical effects that must be accounted for: the reduction of spin structures after polarization sums have been taken over internal massive states.
For example, in the spin-1/2 representation only one factor of the spin vector can appear between a pair of spinors.
This leads to the relations
\begin{subequations}\label{eq:Spin1/2Reduction}
\begin{align}
    \hat{q}_{4,1\mu}\hat{q}_{3,1\nu}\langle\bbv|a^{\mu}_{1/2}|v_{I}\rangle[ v^{I}|a^{\nu}_{1/2}|\bv]
    &=\frac{i}{2m^{2}}\hat{q}_{3,1\mu}\hat{q}_{4,1\nu}\langle\bbv|S^{\mu\nu}_{1/2}|\bv\rangle+\cO(\hbar^{2}), \\
    \hat{q}_{3,2\mu}\hat{q}_{4,2\nu}[\bbv|a^{\mu}_{1/2}|v_{I}]\langle v^{I}|a^{\nu}_{1/2}|\bv\rangle
    &=\frac{i}{2m^{2}}\hat{q}_{4,2\mu}\hat{q}_{3,2\nu}[\bbv|S^{\mu\nu}_{1/2}|\bv]+\cO(\hbar^{2}),
\end{align}
\end{subequations}
for spin-1/2 external states.
By expressing the three-point amplitudes of the previous section using spin vectors and tensors in the spin-1/2 representation, such reductions are easy to perform for particles of any spin.
See \Cref{app:SpinProperties} for more details.

To identify a difference between the black-hole computation and that for a general object, let us examine the individual factorization channels.
Accounting for all $\cO(\hbar\times1/\hbar)$ effects, the factorization channels in the infinite-spin limit take the compact forms\footnote{We can take the infinite-spin limit and drop the overall spinor contraction $\langle\bbv\bv\rangle^{2s}$ now because we have evaluated the polarization sums.}
\begin{subequations}\label{eq:OppositeHelicityChannels}
\begin{align}
    \cF_{1}^{\infty,-+}&=-\frac{y^{4}}{t_{13}s_{34}^{2}}\sum_{j=0}^{\infty}\sum_{k=0}^{\infty}\frac{\left(\hat{q}_{4}\cdot\mathfrak{a}\right)^{j}\left(-\hat{q}_{3}\cdot\mathfrak{a}\right)^{k}}{j!k!}\notag \\
    &\qquad\times\left[C_{S^{k}}C_{S^{j}}-\left(C_{S^{j}}-C_{S^{j+1}}\right)\left(C_{S^{k}}-C_{S^{k+1}}\right)\frac{s_{34}}{2y}w\cdot\mathfrak{a}\right], \\
    \cF_{2}^{\infty,-+}&=-\frac{y^{4}}{t_{14}s_{34}^{2}}\sum_{j=0}^{\infty}\sum_{k=0}^{\infty}\frac{\left(\hat{q}_{4}\cdot\mathfrak{a}\right)^{j}\left(-\hat{q}_{3}\cdot\mathfrak{a}\right)^{k}}{j!k!}\notag \\
    &\qquad\times\left[C_{S^{k}}C_{S^{j}}+\left(C_{S^{j}}-C_{S^{j+1}}\right)\left(C_{S^{k}}-C_{S^{k+1}}\right)\frac{s_{34}}{2y}w\cdot\mathfrak{a}\right].
\end{align}
\end{subequations}
We see in these equations the proof of our statement that BCFW recursion and the classical limit do not commute.
The second terms in the square brackets above are subleading in $\hbar$ relative to the first terms, and are the $\cO(\hbar\times1/\hbar)$ terms we have kept track of.
Since they have different signs in both factorization channels, they do not drop by one power of $\hbar$ when both channels are added, unlike the first terms.
Hence these subleading terms in $\cF_{i}$ are not subleading in the amplitude.
Furthermore, in the black-hole limit $C_{S^{j}}=1$, these contributions vanish and the $\cF_{i}$ are uniform in $\hbar$, demonstrating the cancellations in the black-hole case alluded to in the introduction.

While the black-hole limit implies that the factorization channels are uniform in $\hbar$, the converse is also true.
Specifically, requiring that the factorization channels are uniform in $\hbar$ imposes either $C_{S^{j}}=0$ or $C_{S^{j}}=C_{S^{0}}$ for all $j$.
However, as $C_{S^{0}}=C_{S^{1}}=1$ for any gravitating object \cite{Levi:2015msa}, the former condition cannot be satisfied.
Thus, the uniformity in $\hbar$ of the factorization channels is equivalent to scattering an object with the linear-in-curvature induced multipoles of a Kerr black hole.

As mentioned above, this non-uniformity in $\hbar$ does not persist once the factorization channels are combined into the amplitude.
We will see hints of the exceptionality of the Kerr-black-hole spin-induced multipoles at the level of the amplitude in the same-helicity case, in a way which will be more reminiscent of the notion of minimal coupling of ref.~\cite{Arkani-Hamed:2017jhn}.

Summing the two factorization channels gives the amplitude:
\begin{align}\label{eq:OppositeHelicityCompton}
    \cM^{\infty,-+}_{4}&=\frac{y^{4}}{t_{13}t_{14}s_{34}}\sum_{j=0}^{\infty}\sum_{k=0}^{\infty}\frac{\left(\hat{q}_{4}\cdot\mathfrak{a}\right)^{j}\left(-\hat{q}_{3}\cdot\mathfrak{a}\right)^{k}}{j!k!}\notag \\
    &\qquad\times\left[C_{S^{k}}C_{S^{j}}+\left(C_{S^{j}}-C_{S^{j+1}}\right)\left(C_{S^{k}}-C_{S^{k+1}}\right)\frac{t_{14}-t_{13}}{2y}w\cdot\mathfrak{a}\right],
\end{align}
valid up to fourth order in spin.
Expanding up to third order in spin, we reproduce the helicity-preserving amplitude of ref.~\cite{Saketh:2022wap}.
Up to fourth order in spin, we agree with the result derived from the action of ref.~\cite{Bern:2022kto} for certain choices of their additional parameters.\footnote{We thank Andres Luna and Fei Teng for sharing unpublished Compton amplitudes.}
In the black-hole limit the second term in square brackets is vanishing, thus recovering the spin-exponential of the Compton amplitude in the form presented in ref.~\cite{Aoude:2020onz}.

As in the black-hole case, unphysical poles in $y$ develop above fourth order in spin.
We can remove them without affecting factorization properties exactly as was done in ref.~\cite{Aoude:2022trd}.
We must first isolate the problematic parts of the amplitude, which can be done by plugging in \cref{eq:UnshiftedHats} and using the binomial theorem.
Doing so and collecting like terms gives
\begin{align}
    \cM^{\infty,-+}_{4}&=\sum_{j=0}^{\infty}\sum_{k=0}^{\infty}\sum_{n_{1}=0}^{j}\sum_{n_{2}=0}^{k}{{j}\choose{n_{1}}}{{k}\choose{n_{2}}}\frac{\left(q_{4}\cdot\mathfrak{a}\right)^{j-n_{1}}\left(-q_{3}\cdot\mathfrak{a}\right)^{k-n_{2}}}{2^{n_{1}+n_{2}}j!k!}\notag \\
    &\qquad\times\left[C_{S^{k}}C_{S^{j}}K_{n_{1}+n_{2}}+\frac{1}{2}\left(C_{S^{j}}-C_{S^{j+1}}\right)\left(C_{S^{k}}-C_{S^{k+1}}\right)K_{n_{1}+n_{2}+1}\right],
\end{align}
where
\begin{align}
    K_{n}\equiv\frac{y^{4}}{t_{13}t_{14}s_{34}}\left(\frac{t_{14}-t_{13}}{y}w\cdot\mathfrak{a}\right)^{n}.
\end{align}
Unphysical poles arise in terms containing $K_{n\geq5}$, and can be removed without affecting factorization properties by replacing
\begin{align}
    K_{n}\rightarrow\bar{K}_{n}&\equiv\begin{cases}
        K_{n}, & n\leq4, \\
        K_{4} L_{n-4}  - K_{3} \mathfrak{s}_{2} L_{n-5}, & n>4,
    \end{cases}
\end{align}
with
\begin{align}
    L_{m} &\equiv \sum_{j=0}^{\floor{m/2}} \binom{m+1}{2j+1} \mathfrak{s}^{m-2j}_{1} (\mathfrak{s}^2_1 - \mathfrak{s}_{2})^{j},\\
    \mathfrak{s}_1 &\equiv (q_3 - q_4 )\cdot \mathfrak{a} ,\quad
	\mathfrak{s}_{2} \equiv -4 (q_3 \cdot \mathfrak{a})(q_4 \cdot \mathfrak{a}) + s_{34} \mathfrak{a}^{2}.\notag
\end{align}
We thus arrive at the final, local result -- modulo contact terms -- for the opposite-helicity Compton amplitude for general objects and all spins:
\begin{align}\label{eq:OppositeHelicityCured}
    \cM^{\infty,-+}_{4}&=\sum_{j=0}^{\infty}\sum_{k=0}^{\infty}\sum_{n_{1}=0}^{j}\sum_{n_{2}=0}^{k}{{j}\choose{n_{1}}}{{k}\choose{n_{2}}}\frac{\left(q_{4}\cdot\mathfrak{a}\right)^{j-n_{1}}\left(-q_{3}\cdot\mathfrak{a}\right)^{k-n_{2}}}{2^{n_{1}+n_{2}}j!k!}\notag \\
    &\qquad\times\left[C_{S^{k}}C_{S^{j}}\bar{K}_{n_{1}+n_{2}}+\frac{1}{2}\left(C_{S^{j}}-C_{S^{j+1}}\right)\left(C_{S^{k}}-C_{S^{k+1}}\right)\bar{K}_{n_{1}+n_{2}+1}\right].
\end{align}
Contact terms will be considered in the next section.

Let us end by briefly commenting on previous attempts to construct the Compton amplitude using BCFW recursion on a classical three-point amplitude, namely refs.~\cite{Chen:2021kxt,Chen:2022yxw}.
It was stated in ref.~\cite{Chen:2021kxt} that the spin dependence of both cuts is the same in the black-hole limit.
While this statement is true, it is the result of cancellations between three quantum $\times$ super-classical effects which we have seen above: 1) $\cO(\hbar)$ parts of the three-point amplitudes; 2) $\cO(\hbar^{2})$ parts of the shifted momenta; 3) the reduction of products of spin vectors after the polarization sum over massive internal states.
The authors of refs.~\cite{Chen:2021kxt,Chen:2022yxw} missed the former effect, and their accounting of the latter two did not produce $\cO(\hbar\times1/\hbar)$ terms.\footnote{We thank Jung-Wook Kim for discussions about this.}
Though inconsequential in the black-hole limit, the lack of such effects renders the results of those previous analyses for general spin-induced multipoles discrepant with refs.~\cite{Bern:2022kto,Saketh:2022wap,Levi:2022dqm}, as well as with our results above.
A concrete example of this disagreement is eq.~(B.27) of ref.~\cite{Chen:2021kxt}, which is missing $C_{S^{2}}^{2}$ contributions at cubic order in spin.

\subsection{Same-helicity Compton scattering}

The same-helicity Compton amplitude is much simpler than its opposite-helicity cousin in the case of black-hole scattering, possessing no unphysical poles at any spin order while also expressible as a spin exponential \cite{Johansson:2019dnu,Aoude:2020onz}.
Ironically, then, in the case of general objects, the computation of the amplitude for this helicity configuration is more involved than for opposite helicities.
An attempt to compute this amplitude using recursive techniques can also be found in ref.~\cite{Chen:2021kxt}.
Apart from missing $\cO(\hbar\times1/\hbar)$ contributions, the result there possesses unphysical poles at quadratic order in spin and above, which are not present in the classical computation of ref.~\cite{Saketh:2022wap}.\footnote{The helicity-reversing amplitude of ref.~\cite{Saketh:2022wap} has spurious singularities above linear order in spin as $\theta\rightarrow\pi$. We have written their amplitude in a manifestly local form in \cref{eq:S&VLocal}. We thank Justin Vines for discussions about this.}
In this section we account for the missing contributions to their amplitude and remove the unphysical poles arising from the BCFW computation.

To evaluate the two-positive configuration we begin with the same shift as in the opposite-helicity case.
Accounting for all $\cO(\hbar\times1/\hbar)$ effects, we find the (non-local) infinite-spin, same-helicity amplitude to be
\begin{align}
    \cM_{4}^{\infty,++}&=\frac{y_{++}^{4}}{t_{13}t_{14}s_{34}}\sum_{j=0}^{\infty}\sum_{k=0}^{\infty}\frac{\left(\hat{q}_{4}\cdot \mathfrak{a}\right)^{j}\left(\hat{q}_{3}\cdot \mathfrak{a}\right)^{k}}{j!k!}\notag \\
    &\qquad\times\left[C_{S^{j}}C_{S^{k}}+(C_{S^{j}}-C_{S^{j+1}})(C_{S^{k}}+C_{S^{k+1}})\frac{t_{14}-t_{13}}{2y}w\cdot\mathfrak{a}\right]+\mathcal{B}.
\end{align}
We have defined $w^{\mu}_{++}\equiv[4|p_{1}\sigma^{\mu}|3]/2m$, so that $y_{++}\equiv2p_{1}\cdot w_{++}=m[43]$.
All $\cO(\hbar\times1/\hbar)$ contributions appear in the second term in square brackets.
Similarly to the opposite-helicity case, we can see that these contributions vanish in the black-hole limit, such that we recover the spin-exponential form of the amplitude in ref.~\cite{Aoude:2020onz}.

The recursive approach taken has missed BCFW boundary terms $\mathcal{B}\neq0$, which is signaled by the development of unphysical singularities in $y$ above linear order in spin and since the amplitude does not have the expected $q_{3}\leftrightarrow q_{4}$ crossing symmetry.
Interestingly, in the black-hole case the BCFW computation produces a local and crossing-symmetric amplitude, so boundary terms are not needed in this case \cite{Johansson:2019dnu}.
In the general case, however, our task has become to determine the appropriate boundary terms to restore both locality and crossing symmetry.
We start with the latter.

The missing crossing symmetry can be seen by noting that $w^{\mu}\rightarrow\bar{w}^{\mu}$ under $q_{3}\leftrightarrow q_{4}$.
The origin of this asymmetry is that the $[3,4\rangle$ BCFW shift we have used does not treat the two gravitons identically.
A remedy to this is to simply average the results of the $[3,4\rangle$ and $[4,3\rangle$ shifts.
Since both shifts produce expressions with the correct factorization properties, the average will also have the appropriate residues on physical poles, with the added benefit of posessing the requisite crossing symmetry.
The result of the averaging is
\begin{align}\label{eq:SameHelicityComptonCrossSym}
    &\cM_{4}^{\infty,++}=\frac{y_{++}^{4}}{2t_{13}t_{14}s_{34}}\sum_{j=0}^{\infty}\sum_{k=0}^{\infty}\frac{1}{j!k!}\left\{C_{S^{j}}C_{S^{k}}\left[\left(\hat{q}_{4}\cdot \mathfrak{a}\right)^{j}\left(\hat{q}_{3}\cdot \mathfrak{a}\right)^{k}+\left(\bar{\hat{q}}_{4}\cdot \mathfrak{a}\right)^{j}\left(\bar{\hat{q}}_{3}\cdot \mathfrak{a}\right)^{k}\right]\right. \\
    &\left.+(C_{S^{j}}-C_{S^{j+1}})(C_{S^{k}}+C_{S^{k+1}})\frac{t_{14}-t_{13}}{2}\left[\frac{w\cdot\mathfrak{a}}{y}\left(\hat{q}_{4}\cdot \mathfrak{a}\right)^{j}\left(\hat{q}_{3}\cdot \mathfrak{a}\right)^{k}-\frac{\bar{w}\cdot\mathfrak{a}}{\bar{y}}\left(\bar{\hat{q}}_{3}\cdot \mathfrak{a}\right)^{j}\left(\bar{\hat{q}}_{4}\cdot \mathfrak{a}\right)^{k}\right]\right\}+\mathcal{B}^{\prime},\notag
\end{align}
where $\mathcal{B}^{\prime}$ are the boundary terms needed to restore locality.
The bar over a symbol represents complex conjugation, and we can see from \cref{eq:UnshiftedHats} that $\hat{q}_{4}\rightarrow\bar{\hat{q}}_{3}$ and $\hat{q}_{3}\rightarrow\bar{\hat{q}}_{4}$ under the swap $q_{3}\leftrightarrow q_{4}$.

Moving on to the restoration of locality, we proceed by introducing non-local contact terms which cancel the poles in $y$ and $\bar{y}$.
Both unphysical poles can be removed simultaneously by introducing contact terms with poles in $y\bar{y}=t_{13}t_{14}-m^{2}s_{34}$.
The quantities $w\cdot\mathfrak{a}/y$ and $\bar{w}\cdot\mathfrak{a}/\bar{y}$ are expressible in terms of this product and $w_{++}\cdot\mathfrak{a}/y_{++}$ using \cref{eq:wTow++}.
The removal of unphysical poles in $y\bar{y}$ at quadratic order in spin is shown explicitly in \Cref{app:LocalityExample}.
Up to cubic order in spin, we find agreement with the helicity-reversing amplitude of ref.~\cite{Saketh:2022wap}, which we write covariantly and in a manifestly-local form in \cref{eq:S&VLocal}.
At quartic order in spin, the same-helicity amplitude is
\begin{align}\label{eq:SameHelSpinFour}
    &\cM_{4}^{\infty,++}|_{\mathfrak{a}^{4}}=\frac{y_{++}^{4}}{t_{13}t_{14}s_{34}}\left\{\frac{1}{4!}[(q_{3}+q_{4})\cdot\mathfrak{a}]^{4}\right. \\
    &+\frac{C_{S^{2}}-1}{2!}\frac{s_{34}}{8}\left[24m^{2}t_{13}t_{14}\left(\frac{w_{++}\cdot\mathfrak{a}}{y_{++}}\right)^{4}+2m^{2}s_{34}\left(\frac{w_{++}\cdot\mathfrak{a}}{y_{++}}\right)^{2}\left[\mathfrak{a}^{2}+4m^{2}\left(\frac{w_{++}\cdot\mathfrak{a}}{y_{++}}\right)^{2}\right]\right.\notag \\
    &\qquad\left.+2\mathfrak{a}^{2}(q_{3}\cdot\mathfrak{a})(q_{4}\cdot\mathfrak{a})+[(q_{3}+q_{4})\cdot\mathfrak{a}](t_{14}-t_{13})\frac{w_{++}\cdot\mathfrak{a}}{y_{++}}\left[\mathfrak{a}^{2}+8m^{2}\left(\frac{w_{++}\cdot\mathfrak{a}}{y_{++}}\right)^{2}\right]\right]\notag \\
    &+\frac{(C_{S^{2}}-1)^{2}}{2!2!}\frac{s_{34}}{8}\left[2\mathfrak{a}^{2}(q_{3}\cdot\mathfrak{a})(q_{4}\cdot\mathfrak{a})+[(q_{3}+q_{4})\cdot\mathfrak{a}](t_{14}-t_{13})\frac{w_{++}\cdot\mathfrak{a}}{y_{++}}\left[\mathfrak{a}^{2}+8m^{2}\left(\frac{w_{++}\cdot\mathfrak{a}}{y_{++}}\right)^{2}\right]\right.\notag \\
    &\qquad\left.-2\left(\frac{w_{++}\cdot\mathfrak{a}}{y_{++}}\right)^{2}\left[(2t_{13}t_{14}-m^{2}s_{34})\mathfrak{a}^{2}-4m^{2}(3t_{13}t_{14}+m^{2}s_{34})\left(\frac{w_{++}\cdot\mathfrak{a}}{y_{++}}\right)^{2}\right]\right]\notag \\
    &+(C_{S^{2}}-1)(C_{S^{3}}-1)\frac{s_{34}}{16}[(q_{3}+q_{4})\cdot\mathfrak{a}](t_{14}-t_{13})\frac{w_{++}\cdot\mathfrak{a}}{y_{++}}\left[\mathfrak{a}^{2}+4m^{2}\left(\frac{w_{++}\cdot\mathfrak{a}}{y_{++}}\right)^{2}\right]\notag \\
    &+\frac{C_{S^{3}}-1}{3!}\frac{s_{34}}{4}\left[-24m^{2}t_{13}t_{14}\left(\frac{w_{++}\cdot\mathfrak{a}}{y_{++}}\right)^{4}\right.\notag \\
    &\qquad\left.+\left[(q_{3}\cdot\mathfrak{a})^{2}+(q_{4}\cdot\mathfrak{a})^{2}+2[(q_{3}+q_{4})\cdot\mathfrak{a}](t_{14}-t_{13})\frac{w_{++}\cdot\mathfrak{a}}{y_{++}}\right]\left[\mathfrak{a}^{2}+4m^{2}\left(\frac{w_{++}\cdot\mathfrak{a}}{y_{++}}\right)^{2}\right]\right]\notag \\   
    &+\frac{C_{S^{4}}-1}{4!}\left[[(q_{3}+q_{4})\cdot\mathfrak{a}]^{4}+6m^{2}s_{34}t_{13}t_{14}\left(\frac{w_{++}\cdot\mathfrak{a}}{y_{++}}\right)^{4}-\frac{9s_{34}^{2}}{2}m^{2}\left(\frac{w_{++}\cdot\mathfrak{a}}{y_{++}}\right)^{2}\left[\mathfrak{a}^{2}+4m^{2}\left(\frac{w_{++}\cdot\mathfrak{a}}{y_{++}}\right)^{2}\right]\right.\notag \\
    &\qquad-s_{34}[(q_{3}\cdot\mathfrak{a})^{2}+(q_{4}\cdot\mathfrak{a})^{2}]\left[\mathfrak{a}^{2}+4m^{2}\left(\frac{w_{++}\cdot\mathfrak{a}}{y_{++}}\right)^{2}\right]-\frac{3s_{34}}{2}(q_{3}\cdot\mathfrak{a})(q_{4}\cdot\mathfrak{a})\left[\mathfrak{a}^{2}-8m^{2}\left(\frac{w_{++}\cdot\mathfrak{a}}{y_{++}}\right)^{2}\right]\notag \\
    &\qquad\left.\left.-\frac{s_{34}}{4}[(q_{3}+q_{4})\cdot\mathfrak{a}](t_{14}-t_{13})\frac{w_{++}\cdot\mathfrak{a}}{y_{++}}\left[5\mathfrak{a}^{2}+56m^{2}\left(\frac{w_{++}\cdot\mathfrak{a}}{y_{++}}\right)^{2}\right]\right]\right\}.\notag
\end{align}
Thanks to the overall $y_{++}^{4}$, this expression is also manifestly local.
The symmetry of the amplitude under $q_{3}\leftrightarrow q_{4}$ appears to be broken by the factors of $t_{14}-t_{13}$, but this is not the case: under this exchange, $y_{++}$ is antisymmetric while $w_{++}\cdot\mathfrak{a}$ is symmetric thanks to the spin-supplementary condition $p\cdot\mathfrak{a}=0$, so the combination $(t_{14}-t_{13})w_{++}\cdot\mathfrak{a}/y_{++}$ is itself crossing-symmetric.
\Cref{eq:SameHelSpinFour} agrees with the amplitude derived from the action of ref.~\cite{Bern:2022kto}, up to contact terms.\footnote{Again, we thank Andres Luna and Fei Teng for sharing unpublished results.}

Since $y_{++}\sim m$, we can see from \cref{eq:SameHelSpinFour} (as well as from \cref{eq:S&VLocal}) that the black-hole amplitude scales as $m^{4}$ in the limit $m\rightarrow0$.
Above linear order in spin, the scaling $w_{++}^{\mu}\sim m^{-1}$ dulls this behavior to $m^{2,0,-2}$, at $\mathcal{O}(\mathfrak{a}^{2,3,4})$ respectively.
The best behavior in the high-energy/massless limit thus emerges in the black-hole case, and at high enough spin orders the black-hole limit is the required for the existence of a non-divergent massless limit.
This is reminiscent of the notion of minimal coupling of ref.~\cite{Arkani-Hamed:2017jhn}.

Above fourth order in spin, further non-localities develop in inverse powers of $y_{++}$, analogously to the opposite-helicity case.
The vanishing Gram determinant for the five four-vectors $\mathfrak{a}^{\mu},\,p_{1}^{\mu},\,q_{3}^{\mu},\,q_{4}^{\mu},\,w^{\mu}_{++}$, which reads
\begin{align}\label{eq:SameHelicityGram}
    4m^{4}s_{34}\left(\frac{w_{++}\cdot\mathfrak{a}}{y_{++}}\right)^{2}=-2m^{2}(t_{14}-t_{13})[(q_{3}+q_{4})\cdot\mathfrak{a}]\frac{w_{++}\cdot\mathfrak{a}}{y_{++}}-m^{2}\mathfrak{s}_{2}+\mathfrak{a}^{2}t_{13}t_{14},
\end{align}
allows us to trade terms with more inverse powers of $y_{++}$ for terms with fewer such powers.
Adding this Gram determinant to our arsenal, we have removed all poles in $y_{++}$ from the same-helicity Compton amplitude at fifth order in spin.
We find again that the black-hole amplitude exhibits a finite massless limit at this spin order, with the generic case scaling as $m^{-2}$ when $m\rightarrow0$.
For brevity, we have relegated the amplitude at the fifth order in spin to the ancillary Mathematica notebook \texttt{SameHelicitySpinFourFive.nb}.
\Cref{eq:SameHelSpinFour} is also included in this notebook for convenience.

\vspace{\baselineskip}

Through the evaluation of the classical Compton amplitude for general spinning objects, we have seen in this section that the construction of classical amplitudes using BCFW recursion generally requires that one keep track of subleading parts of intermediate expressions, including quantum pieces of lower-point amplitudes.
Doing so, we have produced for the first time an opposite-helicity Compton amplitude which describes all spin multipoles of a general compact object, which also matches classical computations at low spins.
In the same-helicity case, we presented a non-local, but crossing-symmetric, form of the amplitude to all spins, and cured non-localities up to fifth order in spin.
These amplitudes are not unique, however, as they can be deformed by contact terms.
For the sake of completeness, let us discuss this now.

\section{Contact terms}\label{sec:Contact}

We can write the amplitudes most generally as
\begin{align}
    \cM^{\infty,h_{1}h_{2}}_{4}+m^{2}\left(\mathcal{C}^{h_{1}h_{2}}_{\text{even}}+\mathcal{C}^{h_{1}h_{2}}_{\text{odd}}+\mathcal{D}^{h_{1}h_{2}}_{\text{even}}+\mathcal{D}^{h_{1}h_{2}}_{\text{odd}}\right),
\end{align}
where $\mathcal{C}^{h_{1}h_{2}}_{\text{even}}$ and $\mathcal{C}^{h_{1}h_{2}}_{\text{odd}}$ are (sums of) conservative contact terms at even and odd spin orders respectively, and $\mathcal{D}^{h_{1}h_{2}}_{\text{even}}$ and $\mathcal{D}^{h_{1}h_{2}}_{\text{odd}}$ account for dissipative effects.
We will explain the distinction between the two below.

At the risk of pedantry, let us clarify that a contact term is the product of a coefficient potentially dependent on the scales in the scattering, and a spin structure which is a pole-free function of the momenta and is a monomial in the spin vector.
Most generally, each spin structure is accompanied by its own coefficient.
In the case of Compton scattering, these coefficients are related -- non-trivially -- to the coefficients of curvature-squared operators in a worldline action, operators which describe tidal and spin-induced multipolar effects at quadratic order in curvature.
See ref.~\cite{Levi:2022rrq} for a definition of tidal versus spin-induced multipolar operators in a worldline theory.

An infinite number of contact terms can deform the Compton amplitudes at each spin order, all of which were classified at zeroeth and linear order in spin in refs.~\cite{Haddad:2020que,Aoude:2020ygw}.
However, only a finite number of them contribute to a fixed spin order up to a given order in Newton's constant.
Here we count and write down all contact term deformations contributing to classical gravitational scattering at the same orders as the $C_{S^{k}}$ -- that is to say, at $\cO(G^{2}\mathfrak{a}^{k}/b^{k+1})$, where $b$ is the impact parameter.\footnote{When finite-size effects are allowed, 2PM contributions scale more generally as $\cO(G^{2}R^{j}\mathfrak{a}^{k}/b^{j+k+1})$. This simplifies to $\cO(G^{2+j}\mathfrak{a}^{k}/b^{j+k+1})$ for black holes, so the set of contact terms we consider here can also be thought of as all contact terms potentially relevant to black-hole scattering at $2$PM.}
This is equivalent to requiring that the coefficients of the contact terms do not have any $\hbar$ dependence.
We will illustrate this in more detail now.

\subsection{Relevant scales}

We are working in a context where we have restored factors of $\hbar$ but left $c=1$.
For a general object we therefore have four relevant scales: Planck's constant $\hbar$, Newton's constant $G$, the mass of the object $m$, and the scale of the object's spatial extent $R$.
For a black hole there is one less scale, since the spatial extent is identified with the Schwarzschild radius, which is related to the other scales through $R_{s}=2Gm$.
With $\hbar$ restored, these scales have the dimensions
\begin{align*}
    [R]=[L],\quad [m]=[M],\quad [\hbar]=[L][M],\quad [G]=\frac{[L]}{[M]},
\end{align*}
where $[L]$ represents dimensions of length and $[M]$ dimensions of mass/momentum/energy.

The coupling-stripped amplitude has dimensions $[M]^{2}$, and is $\cO(\hbar^{0})$ in the classical limit.
These dictate the possible scalings of coefficients for classically-relevant contact terms.
We now argue that these properties of the amplitude, combined with the available scales, imply that the contact terms contributing at $\cO(G^{2}\mathfrak{a}^{k}/b^{k+1})$ are those with Wilson coefficients that do not scale with $\hbar$.

Terms in the 2PM scattering angle which scale as $\cO(G^{2}\mathfrak{a}^{k}/b^{k+1})$ in impact-parameter space come from terms of the schematic form $G^{2}q^{k}\mathfrak{a}^{k}/\sqrt{-q^{2}}$ in momentum space, where $q$ is the transfer momentum.
This can be seen to all spin orders in the results of ref.~\cite{Aoude:2022thd}.
The square-root comes from triangle integrals, while the $q^{k}\mathfrak{a}^{k}\sim\cO(\hbar^{0})$ come from $\cO(\hbar^{0})$ parts of the Compton amplitude.
If we consider contact terms in the Compton amplitude with spin structures that scale with some positive power of $\hbar$, the corresponding coefficient must carry a compensating number of inverse factors of $\hbar$ in order for the contact term to scale classically, as was seen in refs.~\cite{Haddad:2020que,Aoude:2020ygw}.
Then, to maintain the correct mass dimensions, the coefficients must also scale with additional powers of $R$ or $Gm$.
This translates to terms of the form $G^{2+n}R^{l}q^{k+j}\mathfrak{a}^{k}/\sqrt{-q^{2}}$ in the one-loop amplitude, where $n+l=j$ is the number of inverse factors of $\hbar$ needed in the contact term coefficient.
Moving to impact-parameter space, these produce terms scaling as $\cO(G^{2+n}R^{l}\mathfrak{a}^{k}/b^{j+k+1})$.

Thus $\cO(G^{2}\mathfrak{a}^{k}/b^{k+1})$ effects are produced by contact terms in the Compton amplitude which have Wilson coefficients that don't depend on $\hbar$ -- or equivalently, spin structures which are $\cO(\hbar^{0})$.
These are the contact terms we will construct in the following.

For spinning objects there is actually an additional scale: the magnitude of the ring radius itself, $|\mathfrak{a}|\equiv\sqrt{\mathfrak{a}^{2}}$.
Allowing the coefficients of their contact terms to depend on the magnitude of the ring radius, the authors of ref.~\cite{Bautista:2022wjf} were able to exactly match the opposite-helicity Compton amplitude to solutions of the Teukolsky equation up to sixth order in spin.
This scale only appeared in their coefficients in the dimensionless combination $\hbar|\mathfrak{a}|/Gm$,\footnote{The scaling of this combination with $\hbar$ is superficial, since in the $\hbar\rightarrow0$ limit the combination $\hbar|\mathfrak{a}|$ is held constant \cite{Maybee:2019jus}.} which does not affect our enumeration of contact terms below since there are no inverse factors of $\hbar$ in this combination that allow us to consider more general spin structures.
The same is true of the other dimensionless combination $\hbar|\mathfrak{a}|/R$, which can appear in the neutron-star case.
More generally, allowing $|\mathfrak{a}|$ to appear independently as a scale in the coefficients is already accounted for by the dissipative contact terms below.

\subsection{Opposite-helicity contact terms}

We would like to explicitly construct all independent contact terms with Wilson coefficients that do not scale with $\hbar$.
Redundancies between contact terms may arise due to the fact that the Gram determinant vanishes for the five four-vectors $\mathfrak{a}^{\mu},\,p_{1}^{\mu},\,q_{3}^{\mu},\,q_{4}^{\mu},\,w^{\mu}$ in four spacetime dimensions.
Employing the vanishing of the Gram determinant,
\begin{align}\label{eq:OppositeHelicityGram}
    (t_{14} - t_{13})^{2} (\wa)^2 = - 4m^2 s_{34} (\wa)^2 + 2 y (t_{14} - t_{13}) \mathfrak{s}_{1} (\wa) - y^2 \mathfrak{s}_{2} ,
\end{align}
these redundancies can be avoided by excluding any contact term containing the left-hand side of \cref{eq:OppositeHelicityGram} as a subfactor.

Now, in order to carry the correct helicity weight, all contact terms must contain exactly four factors of the helicity vector $w^{\mu}$.
Since $w^{\mu}$ is orthogonal to both $q_{3}^{\mu}$ and $q_{4}^{\mu}$, it can only be contracted with $p_{1}^{\mu}$ and $\mathfrak{a}^{\mu}$.
All contact terms must therefore contain a factor of the form
\begin{align}\label{eq:OppositeHelicityCore}
    y^{n}(w\cdot\mathfrak{a})^{4-n},\quad 0\leq n\leq4.
\end{align}
At each $n$ a core factor can be identified that is $\cO(\hbar^{0})$, out of which all contact terms of interest to us can be constructed by multiplying by the following $\cO(\hbar^{0})$ factors:\footnote{One could expand this list by including dressing factors with apparent singularities in $s_{34}$ but whose residues at $s_{34}=0$ actually vanish, as was done in ref.~\cite{Bautista:2022wjf}. However, such terms are redundant in our case as we've instead allowed for factors of $\mathfrak{a}^{2}$ to appear in contact terms. This amounts to a different choice of basis on account of \cref{eq:OppositeHelicityGram}, so we must agree on the total number of free coefficients.}
\begin{align}\label{eq:ContactDressingFactors}
    q_{3}\cdot\mathfrak{a},\quad q_{4}\cdot\mathfrak{a},\quad s_{34}\mathfrak{a}^{2},\quad (t_{14}-t_{13})^{2}\mathfrak{a}^{2}.
\end{align}
For $n\leq2$, the last term in this list need not be considered because of \cref{eq:OppositeHelicityGram}.

At each $n$ one can construct two core factors: one involving a single factor of $|\mathfrak{a}|\equiv\sqrt{\mathfrak{a}^{2}}$, and one without.
Contact terms containing such a factor were argued in ref.~\cite{Bautista:2022wjf} to encode dissipative effects, because their coefficients depend on the boundary conditions at the black hole horizon chosen for solving the Teukolsky equation.
We begin by focusing on conservative contact terms, and subsequently consider dissipative ones.

\subsubsection{Conservative contact terms}

Counting all possible on-shell contact terms is made easier by considering each value of $n$ individually in \cref{eq:OppositeHelicityCore}.
Let us illustrate this counting for $n=4$ and $n=3$.

\vspace{\baselineskip}

\noindent$\boldsymbol{n=4}:$

\vspace{\baselineskip}

The core factor scaling as $\cO(\hbar^{0})$ and carrying the correct helicity weights is $y^{4}\mathfrak{a}^{4}$.
There are no redundancies because of \cref{eq:OppositeHelicityGram} in this case, so we can dress this with any of the factors in \cref{eq:ContactDressingFactors}.

To reach $\cO(\mathfrak{a}^{2k\geq4})$ we must dress the core structure with $2k-4$ powers of spin, with $i$ dressing factors being quadratic in spin.
Of these $i$ we take $l$ factors to be $(t_{14}-t_{13})^{2}\mathfrak{a}^{2}$, and the remaining $i-l$ to be $s_{34}\mathfrak{a}^{2}$.
We are then left with $2k-4-2i$ linear-in-spin dressing factors, of which $j$ are, say, $q_{3}\cdot\mathfrak{a}$.
The total number of even-in-spin contact terms with $n=4$ is then given by the triple sum
\begin{align}\label{eq:Contactn4Even}
    \sum_{i=0}^{k-2}\sum_{l=0}^{i}\sum_{j=0}^{2k-4-2i}1=\frac{1}{6}k(k-1)(2k-1).
\end{align}
The same logic for $\cO(\mathfrak{a}^{2k+1\geq5})$ gives
\begin{align}\label{eq:Contactn4Odd}
    \sum_{i=0}^{k-2}\sum_{l=0}^{i}\sum_{j=0}^{2k-3-2i}1=\frac{1}{3}k(k^{2}-1)
\end{align}
total contact terms.

\vspace{\baselineskip}

\noindent$\boldsymbol{n=3}:$

\vspace{\baselineskip}

The core factor scaling as $\cO(\hbar^{0})$ and carrying the correct helicity weights is $(t_{14}-t_{13})y^{3}\mathfrak{a}^{4}(w\cdot\mathfrak{a})$.
Again, we cannot have redundancies due to the vanishing Gram determinant in this case.

Since the core factor already has five spin powers, contact terms for $n=3$ only arise at even-in-spin orders from $\cO(\mathfrak{a}^{2k\geq6})$.
This changes the upper bounds on the sums over $i$ and $j$ in \cref{eq:Contactn4Even}, since now quadratic-in-spin dressings can only begin to appear for $k\geq4$ and since five instead of four powers of spin are accounted for in the core factor.
Making these modifications, the total number of even-in-spin contact terms for $n=3$ is
\begin{align}\label{eq:Contactn3Even}
    \sum_{i=0}^{k-3}\sum_{l=0}^{i}\sum_{j=0}^{2k-5-2i}1=\frac{1}{3}k(k-1)(k-2).
\end{align}
For $\cO(\mathfrak{a}^{2k+1\geq5})$ we have
\begin{align}\label{eq:Contactn3Odd}
    \sum_{i=0}^{k-2}\sum_{l=0}^{i}\sum_{j=0}^{2k-4-2i}1=\frac{1}{6}k(k-1)(2k-1)
\end{align}
total contact terms.

\vspace{\baselineskip}

The core factors for the remaining values of $n$ are
\begin{align*}
    n=2:&\quad y^{2}\mathfrak{a}^{2}(w\cdot\mathfrak{a})^{2}, \\
    n=1:&\quad y(t_{14}-t_{13})\mathfrak{a}^{2}(w\cdot\mathfrak{a})^{3}, \\
    n=0:&\quad (w\cdot\mathfrak{a})^{4}.
\end{align*}
In each of these cases we must not dress the core factors with $(t_{14}-t_{13})^{2}\mathfrak{a}^{2}$, as such dressings are reducible using \cref{eq:OppositeHelicityGram}.
Consequently, the counting of contact terms for $n=2,0$ is the same as for the $n=4$ case but with $l$ fixed to zero.
Similarly, the counting for $n=1$ is given by the counting for $n=3$ with the sum over $l$ dropped.
The result of the counting of independent coefficients is given in \cref{tab:NumberEvenContactTerms,tab:NumberOddContactTerms}.

The most-general, conservative, contact-term deformation of the opposite-helicity amplitude relevant at $\cO(G^{2}\mathfrak{a}^{2k}/b^{2k+1})$ is
\begin{align}
    &\mathcal{C}_{\text{even}}^{-+}=\sum_{k=0}^{\infty}\left\{\sum_{i=0}^{k-2}\sum_{j=0}^{2k-4-2i}(s_{34}\mathfrak{a}^{2})^{i}(q_{3}\cdot\mathfrak{a})^{j}(q_{4}\cdot\mathfrak{a})^{2k-4-2i-j}\right.\notag \\
    &\times\left[y^{4}\mathfrak{a}^{4}\sum_{l=0}^{i}a^{-+,k}_{i,l,j}(s_{34}\mathfrak{a}^{2})^{-l}[(t_{14}-t_{13})^{2}\mathfrak{a}^{2}]^{l}+(w\cdot\mathfrak{a})^{2}\left[y^{2}\mathfrak{a}^{2}c^{-+,k}_{i,j}+(w\cdot\mathfrak{a})^{2}e^{-+,k}_{i,j}\right]\right]\notag \\
    &+(t_{14}-t_{13})y\mathfrak{a}^{2}(w\cdot\mathfrak{a})\sum_{i=0}^{k-3}\sum_{j=0}^{2k-5-2i}(s_{34}\mathfrak{a}^{2})^{i}(q_{3}\cdot\mathfrak{a})^{j}(q_{4}\cdot\mathfrak{a})^{2k-5-2i-j}\notag \\
    &\left.\times\left[y^{2}\mathfrak{a}^{2}\sum_{l=0}^{i}b^{-+,k}_{i,l,j}(s_{34}\mathfrak{a}^{2})^{-l}[(t_{14}-t_{13})^{2}\mathfrak{a}^{2}]^{l}+d^{-+,k}_{i,j}(w\cdot\mathfrak{a})^{2}\right]\right\},
\end{align}
where the coefficients $a,\,b,\,c,\,d,\,e$ represent contact terms with $n=4,3,2,1,0$ respectively in \cref{eq:OppositeHelicityCore}.
At $\cO(G^{2}\mathfrak{a}^{2k+1}/b^{2k+2})$ the most-general set of conservative contact terms is
\begin{align}
    &\mathcal{C}_{\text{odd}}^{-+}=\sum_{k=0}^{\infty}\left\{\sum_{i=0}^{k-2}\sum_{j=0}^{2k-3-2i}(s_{34}\mathfrak{a}^{2})^{i}(q_{3}\cdot\mathfrak{a})^{j}(q_{4}\cdot\mathfrak{a})^{2k-3-2i-j}\right.\notag \\
    &\times\left[y^{4}\mathfrak{a}^{4}\sum_{l=0}^{i}\tilde{a}^{-+,k}_{i,l,j}(s_{34}\mathfrak{a}^{2})^{-l}[(t_{14}-t_{13})^{2}\mathfrak{a}^{2}]^{l}+(w\cdot\mathfrak{a})^{2}\left[y^{2}\mathfrak{a}^{2}\tilde{c}^{-+,k}_{i,j}+(w\cdot\mathfrak{a})^{2}\tilde{e}^{-+,k}_{i,j}\right]\right]\notag \\
    &+(t_{14}-t_{13})y\mathfrak{a}^{2}(w\cdot\mathfrak{a})\sum_{i=0}^{k-2}\sum_{j=0}^{2k-4-2i}(s_{34}\mathfrak{a}^{2})^{i}(q_{3}\cdot\mathfrak{a})^{j}(q_{4}\cdot\mathfrak{a})^{2k-4-2i-j}\notag \\
    &\left.\times\left[y^{2}\mathfrak{a}^{2}\sum_{l=0}^{i}\tilde{b}^{-+,k}_{i,l,j}(s_{34}\mathfrak{a}^{2})^{-l}[(t_{14}-t_{13})^{2}\mathfrak{a}^{2}]^{l}+\tilde{d}^{-+,k}_{i,j}(w\cdot\mathfrak{a})^{2}\right]\right\},
\end{align}
All spin structures in the above are $\cO(\hbar^{0})$, so the coefficients for classical contributions must be $\hbar$-free.
However, spin structures containing subfactors of $(t_{14}-t_{13})^{2}\mathfrak{a}^{2}$ have mass dimensions which must be compensated by their coefficients.
Specifically,
\begin{align*}
    [a^{-+,k}_{i,l,j}]=[\tilde{a}^{-+,k}_{i,l,j}]&=[b^{-+,k}_{i,l,j}]=[\tilde{b}^{-+,k}_{i,l,j}]=[M]^{-4-2l}, \\
    [c^{-+,k}_{i,j}]=[\tilde{c}^{-+,k}_{i,j}]&=[d^{-+,k}_{i,j}]=[\tilde{d}^{-+,k}_{i,j}]=[M]^{-2}.
\end{align*}
To achieve this without altering the $\hbar$ scaling of the contact terms, we set\footnote{Other possible combinations of the relevant scales that can produce the requisite mass dimensions are powers of $R/\hbar$, $Gm/\hbar$, or $G/R$. The first two are not classically relevant, while the last one is only relevant for neutron stars past 2PM.}
\begin{align*}
    \{a^{-+,k}_{i,l,j},\tilde{a}^{-+,k}_{i,l,j},b^{-+,k}_{i,l,j},\tilde{b}^{-+,k}_{i,l,j}\}&\propto m^{-4-2l}, \\
    \{c^{-+,k}_{i,j},\tilde{c}^{-+,k}_{i,j},d^{-+,k}_{i,j},\tilde{d}^{-+,k}_{i,j}\}&\propto m^{-2}.
\end{align*}
All other coefficients must be either scaleless or depend on (products of) the factors $\hbar|\mathfrak{a}|/Gm$ and $\hbar|\mathfrak{a}|/R$.

\subsubsection{Dissipative contact terms}

To count dissipative contact terms we modify slightly the core factors to carry the correct helicity weights, scale as $\cO(\hbar^{0})$, and also possess one factor of $|\mathfrak{a}|$ \cite{Bautista:2022wjf}:
\begin{align*}
    n=4:&\quad y^{4}\mathfrak{a}^{4}(t_{14}-t_{13})|\mathfrak{a}|, \\
    n=3:&\quad y^{3}\mathfrak{a}^{2}(w\cdot\mathfrak{a})|\mathfrak{a}|, \\
    n=2:&\quad y^{2}\mathfrak{a}^{2}(w\cdot\mathfrak{a})^{2}(t_{14}-t_{13})|\mathfrak{a}|, \\
    n=1:&\quad y(w\cdot\mathfrak{a})^{3}|\mathfrak{a}|\\
    n=0:&\quad (w\cdot\mathfrak{a})^{4}(t_{14}-t_{13})|\mathfrak{a}|.
\end{align*}
All contact terms can again be generated by dressing these core factors with the dressing factors in \cref{eq:ContactDressingFactors}.
At $\cO(\mathfrak{a}^{4,5})$ ref.~\cite{Bautista:2022wjf} only included dissipative contact terms with factors of $(t_{14}-t_{13})|\mathfrak{a}|$ in their ansatz, while here we also include terms of the form $y|\mathfrak{a}|$.
In ref.~\cite{Bautista:2022wjf} the latter were found to not be needed to match to solutions of the Teukolsky equation.
So, to compare our counting of dissipative contact terms to that in ref.~\cite{Bautista:2022wjf} at these spin orders, we must ignore those emerging from the $n=1,3$ core factors.
At $\cO(\mathfrak{a}^{6})$ ref.~\cite{Bautista:2022wjf} included some terms with these core factors.\footnote{Specifically, the contact terms there with the $c_{10}^{(i)}$ coefficients can be rewritten using \cref{eq:OppositeHelicityGram} to involve terms with the $n=1,3$ core factors. We thank Yilber Fabian Bautista for discussions about this.}

Counting in an identical fashion to the conservative case, we find the number of dissipative contact terms in \cref{tab:NumberEvenDissContactTerms,tab:NumberOddDissContactTerms} for even and odd spin powers respectively.

The most-general set of dissipative contact terms relevant at $\cO(G^{2}\mathfrak{a}^{2k}/b^{2k+1})$ is \begin{align}
    &\mathcal{D}^{-+}_{\text{even}}=|\mathfrak{a}|\sum_{k=0}^{\infty}\left\{(t_{14}-t_{13})\sum_{i=0}^{k-3}\sum_{j=0}^{2k-2i-5}(s_{34}\mathfrak{a}^{2})^{i}(q_{3}\cdot\mathfrak{a})^{j}(q_{4}\cdot\mathfrak{a})^{2k-2i-5-j}\right.\notag \\
    &\times\left[y^{4}\mathfrak{a}^{4}\sum_{l=0}^{i}f^{-+,k}_{i,l,j}(s_{34}\mathfrak{a}^{2})^{-l}[(t_{14}-t_{13})^{2}\mathfrak{a}^{2}]^{l}+(w\cdot\mathfrak{a})^{2}\left[p^{-+,k}_{i,j}y^{2}\mathfrak{a}^{2}+r^{-+,k}_{i,j}(w\cdot\mathfrak{a})^{2}\right]\right]\notag \\
    &+y(w\cdot\mathfrak{a})\sum_{i=0}^{k-2}\sum_{j=0}^{2k-2i-4}(s_{34}\mathfrak{a}^{2})^{i}(q_{3}\cdot\mathfrak{a})^{j}(q_{4}\cdot\mathfrak{a})^{2k-2i-4-j}\notag \\
    &\left.\times\left[y^{2}\mathfrak{a}^{2}\sum_{l=0}^{i}g^{-+,k}_{i,l,j}(s_{34}\mathfrak{a}^{2})^{-l}[(t_{14}-t_{13})^{2}\mathfrak{a}^{2}]^{l}+q^{-+,k}_{i,j}(w\cdot\mathfrak{a})^{2}\right]\right\}.
\end{align}
The coefficients labelled $f,\,g,\,p,\,q,\,r$ correspond to the $n=4,3,2,1,0$ core factors, respectively.
The dissipative contact terms needed at $\cO(G^{2}\mathfrak{a}^{2k+1}/b^{2k+2})$ are 
\begin{align}
    &\mathcal{D}^{-+}_{\text{odd}}=|\mathfrak{a}|\sum_{k=0}^{\infty}\left\{(t_{14}-t_{13})\sum_{i=0}^{k-2}\sum_{j=0}^{2k-2i-4}(s_{34}\mathfrak{a}^{2})^{i}(q_{3}\cdot\mathfrak{a})^{j}(q_{4}\cdot\mathfrak{a})^{2k-2i-4-j}\right.\notag \\
    &\times\left[y^{4}\mathfrak{a}^{4}\sum_{l=0}^{i}\tilde{f}^{-+,k}_{i,l,j}(s_{34}\mathfrak{a}^{2})^{-l}[(t_{14}-t_{13})^{2}\mathfrak{a}^{2}]^{l}+(w\cdot\mathfrak{a})^{2}\left[\tilde{p}^{-+,k}_{i,j}y^{2}\mathfrak{a}^{2}+\tilde{r}^{-+,k}_{i,j}(w\cdot\mathfrak{a})^{2}\right]\right]\notag \\
    &+y(w\cdot\mathfrak{a})\sum_{i=0}^{k-2}\sum_{j=0}^{2k-2i-3}(s_{34}\mathfrak{a}^{2})^{i}(q_{3}\cdot\mathfrak{a})^{j}(q_{4}\cdot\mathfrak{a})^{2k-2i-3-j}\notag \\
    &\left.\times\left[y^{2}\mathfrak{a}^{2}\sum_{l=0}^{i}\tilde{g}^{-+,k}_{i,l,j}(s_{34}\mathfrak{a}^{2})^{-l}[(t_{14}-t_{13})^{2}\mathfrak{a}^{2}]^{l}+\tilde{q}^{-+,k}_{i,j}(w\cdot\mathfrak{a})^{2}\right]\right\}.
\end{align}
As in the conservative case, the condition that the amplitude has mass dimension 2 and scales as $\cO(\hbar^{0})$ in the classical limit imposes certain scalings on the parameters.
In this case, we must have
\begin{align*}
    \{f^{-+,k}_{i,l,j},\tilde{f}^{-+,k}_{i,l,j}\}&\propto m^{-5-2l}, \\
    \{g^{-+,k}_{i,l,j},\tilde{g}^{-+,k}_{i,l,j}\}&\propto m^{-3-2l}, \\
    \{p^{-+,k}_{i,j},\tilde{p}^{-+,k}_{i,j}\}&\propto m^{-3}, \\
    \{q^{-+,k}_{i,j},\tilde{q}^{-+,k}_{i,j},r^{-+,k}_{i,j},\tilde{r}^{-+,k}_{i,j}\}&\propto m^{-1}.
\end{align*}

Imposing crossing symmetry on the scattering renders nearly half of the parameters redundant.
Both opposite-helicity configurations are related under crossing through
\begin{align}\label{eq:OppositeHelicityCrossing}
    \cM_{4}^{+-}=\cM_{4}^{-+}|_{q_{3}\leftrightarrow q_{4}}=\bar{\cM}_{4}^{-+}|_{\mathfrak{a}\rightarrow-\mathfrak{a}}.
\end{align}
Thus the coefficients of contact terms with the subfactor $(q_{3}\cdot\mathfrak{a})^{i}(q_{4}\cdot\mathfrak{a})^{j}$ are related to those of the analogous contact terms with $i$ and $j$ flipped.
For example, two such related coefficients at $\cO(\mathfrak{a}^{5})$ are
\begin{align*}
    \tilde{e}^{-+,2}_{0,0}(w\cdot\mathfrak{a})^{4} q_{4}\cdot\mathfrak{a},\quad \tilde{e}^{-+,2}_{0,1}(w\cdot\mathfrak{a})^{4} q_{3}\cdot\mathfrak{a},
\end{align*}
and \cref{eq:OppositeHelicityCrossing} imposes $\tilde{e}^{-+,2}_{0,0}=-\tilde{e}^{-+,2}_{0,1}$.
The numbers of independent coefficients consistent with crossing symmetry are also shown in \cref{tab:NumberEvenContactTerms,tab:NumberOddContactTerms,tab:NumberEvenDissContactTerms,tab:NumberOddDissContactTerms}.
The number of crossing-symmetric contact terms agrees with ref.~\cite{Bautista:2022wjf} for the conservative sector, while we have additional terms in the dissipative sector which they found to be unnecessary for matching to solutions of the Teukolsky equation.

\subsection{Same-helicity contact terms}

Moving on to same-helicity scattering, the analysis is nearly identical to the opposite-helicity scenario, with the primary difference being that we now work with the helicity vector $w_{++}^{\mu}$ defined above instead of $w^{\mu}$.
Redundancies must again be accounted for because of the vanishing of the Gram determinant \cref{eq:SameHelicityGram}.
It is thus sufficient to construct contact terms that do not include the left-hand side of \cref{eq:SameHelicityGram} as a subfactor.

Analogously to the opposite-helicity case, each contact term must contain four powers of the helicity vector $w_{++}^{\mu}$ in order to transform appropriately under the little groups of the external massless particles.
In this case, the helicity vector is orthogonal to $q_{3}^{\mu}$ but not to $q_{4}^{\mu}$.
Nevertheless, it is easy to show that $q_{4}\cdot w_{++}=-\frac{t_{14}}{2m^{2}}y_{++}$, so this contraction may be ignored so long as we account for $y_{++}$.
Therefore, each contact term must contain a factor of
\begin{align}\label{eq:SameHelicityCore}
    y_{++}^{n}(w_{++}\cdot\mathfrak{a})^{4-n},\quad 0\leq n\leq4.
\end{align}
Another option exists for the helicity vector: $\tilde{w}_{++}^{\mu}=[4|\gamma^{\mu}p_{1}|3]/2m$.
We can work with $w_{++}^{\mu}$ exclusively since the two are related by $\tilde{w}_{++}^{\mu}=\frac{p_{1}^{\mu}}{m^{2}}y_{++}-w_{++}^{\mu}$, and hence any contraction with $\tilde{w}_{++}^{\mu}$ is already accounted for in terms of contractions with $w_{++}^{\mu}$.

At a fixed $n$ the core factors for both conservative and dissipative contact terms are identical to the opposite-helicity core factors, but with $\{y,w^{\mu}\}\rightarrow\{y_{++},w_{++}^{\mu}\}$.
Contact terms are then made by dressing the core factors with the factors in \cref{eq:ContactDressingFactors}.
Now, however, \cref{eq:SameHelicityGram} tells us that we must not use the third of these dressing factors for $n\leq2$.

All-in-all, carrying out the counting as above shows that there is the same number of conservative and dissipative contact terms at each fixed $n$ in both the general and crossing-symmetric sectors as for opposite-helicity scattering; see \cref{tab:NumberEvenContactTerms,tab:NumberOddContactTerms,tab:NumberEvenDissContactTerms,tab:NumberOddDissContactTerms}.
The forms of the contact terms are slightly different, however, because of the differing Gram determinants between both helicity configuations.

\subsubsection{Conservative contact terms}

The most-general, conservative, contact-term deformation of the same-helicity amplitude relevant at $\cO(G^{2}\mathfrak{a}^{2k}/b^{2k+1})$ is
\begin{align}
    &\mathcal{C}_{\text{even}}^{++}=\sum_{k=0}^{\infty}\left\{\sum_{i=0}^{k-2}\sum_{j=0}^{2k-4-2i}[(t_{14}-t_{13})^{2}\mathfrak{a}^{2}]^{i}(q_{3}\cdot\mathfrak{a})^{j}(q_{4}\cdot\mathfrak{a})^{2k-4-2i-j}\right.\notag \\
    &\times\left[y_{++}^{4}\mathfrak{a}^{4}\sum_{l=0}^{i}a^{++,k}_{i,l,j}(s_{34}\mathfrak{a}^{2})^{l}[(t_{14}-t_{13})^{2}\mathfrak{a}^{2}]^{-l}+(w_{++}\cdot\mathfrak{a})^{2}\left[y_{++}^{2}\mathfrak{a}^{2}c^{++,k}_{i,j}+(w_{++}\cdot\mathfrak{a})^{2}e^{++,k}_{i,j}\right]\right]\notag \\
    &+(t_{14}-t_{13})y_{++}\mathfrak{a}^{2}(w_{++}\cdot\mathfrak{a})\sum_{i=0}^{k-3}\sum_{j=0}^{2k-5-2i}[(t_{14}-t_{13})^{2}\mathfrak{a}^{2}]^{i}(q_{3}\cdot\mathfrak{a})^{j}(q_{4}\cdot\mathfrak{a})^{2k-5-2i-j}\notag \\
    &\left.\times\left[y^{2}_{++}\mathfrak{a}^{2}\sum_{l=0}^{i}b^{++,k}_{i,l,j}(s_{34}\mathfrak{a}^{2})^{l}[(t_{14}-t_{13})^{2}\mathfrak{a}^{2}]^{-l}+d^{++,k}_{i,j}(w_{++}\cdot\mathfrak{a})^{2}\right]\right\}.
\end{align}
At $\cO(G^{2}\mathfrak{a}^{2k+1}/b^{2k+2})$ the set is
\begin{align}
    &\mathcal{C}^{++}_{\text{odd}}=\sum_{k=0}^{\infty}\left\{\sum_{i=0}^{k-2}\sum_{j=0}^{2k-3-2i}[(t_{14}-t_{13})^{2}\mathfrak{a}^{2}]^{i}(q_{3}\cdot\mathfrak{a})^{j}(q_{4}\cdot\mathfrak{a})^{2k-3-2i-j}\right.\notag \\
    &\times\left[y_{++}^{4}\mathfrak{a}^{4}\sum_{l=0}^{i}\tilde{a}^{++,k}_{i,l,j}(s_{34}\mathfrak{a}^{2})^{l}[(t_{14}-t_{13})^{2}\mathfrak{a}^{2}]^{-l}+(w_{++}\cdot\mathfrak{a})^{2}\left[y_{++}^{2}\mathfrak{a}^{2}\tilde{c}^{++,k}_{i,j}+(w_{++}\cdot\mathfrak{a})^{2}\tilde{e}^{++,k}_{i,j}\right]\right]\notag \\
    &+(t_{14}-t_{13})y_{++}\mathfrak{a}^{2}(w_{++}\cdot\mathfrak{a})\sum_{i=0}^{k-2}\sum_{j=0}^{2k-4-2i}[(t_{14}-t_{13})^{2}\mathfrak{a}^{2}]^{i}(q_{3}\cdot\mathfrak{a})^{j}(q_{4}\cdot\mathfrak{a})^{2k-4-2i-j}\notag \\
    &\left.\times\left[y^{2}_{++}\mathfrak{a}^{2}\sum_{l=0}^{i}\tilde{b}^{++,k}_{i,l,j}(s_{34}\mathfrak{a}^{2})^{l}[(t_{14}-t_{13})^{2}\mathfrak{a}^{2}]^{-l}+\tilde{d}^{++,k}_{i,j}(w_{++}\cdot\mathfrak{a})^{2}\right]\right\}.
\end{align}
As above, contributions with a classical $\hbar$ scaling require 
\begin{align*}
    \{a^{++,k}_{i,l,j},\tilde{a}^{++,k}_{i,l,j},b^{++,k}_{i,l,j},\tilde{b}^{++,k}_{i,l,j}\}&\propto m^{-4-2(i-l)}, \\
    \{c^{++,k}_{i,j},\tilde{c}^{++,k}_{i,j},d^{++,k}_{i,j},\tilde{d}^{++,k}_{i,j}\}&\propto m^{-2-2i}, \\
    \{e^{++,k}_{i,j},\tilde{e}^{++,k}_{i,j}\}&\propto m^{-2i}.
\end{align*}
Unlike in the conservative sector of the opposite-helicity amplitude, here all parameters must scale with some power of the mass unless $i=0$.

\subsubsection{Dissipative contact terms}

The most-general set of dissipative contact terms relevant at $\cO(G^{2}\mathfrak{a}^{2k}/b^{2k+1})$ is \begin{align}
    &\mathcal{D}^{++}_{\text{even}}=|\mathfrak{a}|\sum_{k=0}^{\infty}\left\{(t_{14}-t_{13})\sum_{i=0}^{k-3}\sum_{j=0}^{2k-2i-5}[(t_{14}-t_{13})^{2}\mathfrak{a}^{2}]^{i}(q_{3}\cdot\mathfrak{a})^{j}(q_{4}\cdot\mathfrak{a})^{2k-2i-5-j}\right.\notag \\
    &\times\left[y^{4}_{++}\mathfrak{a}^{4}\sum_{l=0}^{i}f^{++,k}_{i,l,j}(s_{34}\mathfrak{a}^{2})^{l}[(t_{14}-t_{13})^{2}\mathfrak{a}^{2}]^{-l}+(w_{++}\cdot\mathfrak{a})^{2}\left[p^{++,k}_{i,l,j}y^{2}_{++}\mathfrak{a}^{2}+r^{++,k}_{i,l,j}(w_{++}\cdot\mathfrak{a})^{2}\right]\right]\notag \\
    &+y_{++}(w_{++}\cdot\mathfrak{a})\sum_{i=0}^{k-2}\sum_{j=0}^{2k-2i-4}[(t_{14}-t_{13})^{2}\mathfrak{a}^{2}]^{i}(q_{3}\cdot\mathfrak{a})^{j}(q_{4}\cdot\mathfrak{a})^{2k-2i-4-j}\notag \\
    &\left.\times\left[y^{2}_{++}\mathfrak{a}^{2}\sum_{l=0}^{i}g^{++,k}_{i,l,j}(s_{34}\mathfrak{a}^{2})^{l}[(t_{14}-t_{13})^{2}\mathfrak{a}^{2}]^{-l}+q^{++,k}_{i,l,j}(w_{++}\cdot\mathfrak{a})^{2}\right]\right\},
\end{align}
while at $\cO(G^{2}\mathfrak{a}^{2k+1}/b^{2k+2})$ we find 
\begin{align}
    &\mathcal{D}^{++}_{\text{odd}}=|\mathfrak{a}|\sum_{k=0}^{\infty}\left\{(t_{14}-t_{13})\sum_{i=0}^{k-2}\sum_{j=0}^{2k-2i-4}[(t_{14}-t_{13})^{2}\mathfrak{a}^{2}]^{i}(q_{3}\cdot\mathfrak{a})^{j}(q_{4}\cdot\mathfrak{a})^{2k-2i-4-j}\right.\notag \\
    &\times\left[y^{4}_{++}\mathfrak{a}^{4}\sum_{l=0}^{i}\tilde{f}^{++,k}_{i,l,j}(s_{34}\mathfrak{a}^{2})^{l}[(t_{14}-t_{13})^{2}\mathfrak{a}^{2}]^{-l}+(w_{++}\cdot\mathfrak{a})^{2}\left[\tilde{p}^{++,k}_{i,l,j}y^{2}_{++}\mathfrak{a}^{2}+\tilde{r}^{++,k}_{i,l,j}(w_{++}\cdot\mathfrak{a})^{2}\right]\right]\notag \\
    &+y_{++}(w_{++}\cdot\mathfrak{a})\sum_{i=0}^{k-2}\sum_{j=0}^{2k-2i-3}[(t_{14}-t_{13})^{2}\mathfrak{a}^{2}]^{i}(q_{3}\cdot\mathfrak{a})^{j}(q_{4}\cdot\mathfrak{a})^{2k-2i-3-j}\notag \\
    &\left.\times\left[y^{2}_{++}\mathfrak{a}^{2}\sum_{l=0}^{i}\tilde{g}^{++,k}_{i,l,j}(s_{34}\mathfrak{a}^{2})^{l}[(t_{14}-t_{13})^{2}\mathfrak{a}^{2}]^{-l}+\tilde{q}^{++,k}_{i,l,j}(w_{++}\cdot\mathfrak{a})^{2}\right]\right\}.
\end{align}
In this final set of contact terms, the coefficients contributing at $\cO(\hbar^{0})$ have the scalings
\begin{align*}
    \{f^{++,k}_{i,l,j},\tilde{f}^{++,k}_{i,l,j}\}&\propto m^{-5-2(i-l)}, \\
    \{g^{-+,k}_{i,l,j},\tilde{g}^{-+,k}_{i,l,j}\}&\propto m^{-3-2(i-l)}, \\
    \{p^{-+,k}_{i,j},\tilde{p}^{-+,k}_{i,j}\}&\propto m^{-3-2i}, \\
    \{q^{-+,k}_{i,j},\tilde{q}^{-+,k}_{i,j},r^{-+,k}_{i,j},\tilde{r}^{-+,k}_{i,j}\}&\propto m^{-1-2i}.
\end{align*}

Crossing symmetry is satisfied if
\begin{align}
    \cM^{--}_{4}=\bar{\cM}_{4}^{++}|_{\mathfrak{a}\rightarrow-\mathfrak{a}},\quad \cM_{4}^{++}=\cM_{4}^{++}|_{q_{3}\leftrightarrow q_{4}}.
\end{align}
The first of these determines the amplitude with two negative-helicity gravitons, while the second constrains many of the free coefficients.
\Cref{tab:NumberEvenContactTerms,tab:NumberOddContactTerms,tab:NumberEvenDissContactTerms,tab:NumberOddDissContactTerms} show the number of remaining free coefficients after requiring crossing symmetry.

\newpage

\begin{table}
    \centering
    \begin{tabular}{c|c|c}
         & general & crossing-symmetric \\
        \hline
        \hline
        $n=4$ & $\frac{1}{6}k(k-1)(2k-1)$ & $\frac{1}{6}k(k^{2}-1)$ \\
        \hline
        $n=3$ & $\frac{1}{3}k(k-1)(k-2)$ & $\frac{1}{6}k(k-1)(k-2)$ \\
        \hline
        $n=2$ & $(k-1)^{2}$ & $\frac{1}{2}k(k-1)$ \\
        \hline
        $n=1$ & $(k-1)(k-2)$ & $\frac{1}{2}(k-1)(k-2)$ \\
        \hline
        $n=0$ & $(k-1)^{2}$ & $\frac{1}{2}k(k-1)$ \\
        \hline
        \hline
        Total & $\frac{1}{6}(k-1)(4k^{2}+13k-24)$ & $\frac{1}{3}(k-1)(k^{2}+4k-3)$
    \end{tabular}
    \caption{The total number of independent $\cO(\mathfrak{a}^{2k\geq4})$ conservative contact term coefficients, and the number left over after requiring crossing symmetry. Valid for all helicity configurations. The number of crossing-symmetric contact terms at $\cO(\mathfrak{a}^{4,6})$ agrees with the counting of ref.~\cite{Bautista:2022wjf} for the opposite-helicity amplitude.}
    \label{tab:NumberEvenContactTerms}
\end{table}
\begin{table}
    \centering
    \begin{tabular}{c|c|c}
         & general & crossing-symmetric \\
        \hline
        \hline
        $n=4$ & $\frac{1}{3}k(k^{2}-1)$ & $\frac{1}{6}k(k^{2}-1)$ \\
        \hline
        $n=3$ & $\frac{1}{6}k(k-1)(2k-1)$ & $\frac{1}{6}k(k^{2}-1)$ \\
        \hline
        $n=2$ & $k(k-1)$ & $\frac{1}{2}k(k-1)$ \\
        \hline
        $n=1$ & $(k-1)^{2}$ & $\frac{1}{2}k(k-1)$ \\
        \hline
        $n=0$ & $k(k-1)$ & $\frac{1}{2}k(k-1)$ \\
        \hline
        \hline
        Total & $\frac{1}{6}(k-1)(4k^{2}+19k-6)$ & $\frac{1}{6}k(k-1)(2k+11)$
    \end{tabular}
    \caption{The total number of independent $\cO(\mathfrak{a}^{2k+1\geq5})$ conservative contact term coefficients, and the number left over after requiring crossing symmetry. Valid for all helicity configurations. The number of crossing-symmetric contact terms at $\cO(\mathfrak{a}^{5})$ agrees with the counting of ref.~\cite{Bautista:2022wjf} for the opposite-helicity amplitude.}
    \label{tab:NumberOddContactTerms}
\end{table}

\begin{table}
    \centering
    \begin{tabular}{c|c|c}
         & general & crossing-symmetric \\
        \hline
        \hline
        $n=4$ & $\frac{1}{3}k(k-1)(k-2)$ & $\frac{1}{6}k(k-1)(k-2)$ \\
        \hline
        $n=3$ & $\frac{1}{6}k(k-1)(2k-1)$ & $\frac{1}{6}k(k^{2}-1)$ \\
        \hline
        $n=2$ & $(k-1)(k-2)$ & $\frac{1}{2}(k-1)(k-2)$ \\
        \hline
        $n=1$ & $(k-1)^{2}$ & $\frac{1}{2}k(k-1)$ \\
        \hline
        $n=0$ & $(k-1)(k-2)$ & $\frac{1}{2}(k-1)(k-2)$ \\
        \hline
        \hline
        Total & $\frac{1}{6}(k-1)(4k^{2}+13k-30)$ & $\frac{1}{3}(k-1)(k^{2}+4k-6)$
    \end{tabular}
    \caption{The total number of independent $\cO(\mathfrak{a}^{2k\geq4})$ dissipative contact term coefficients, and the number left over after requiring crossing symmetry. Valid for all helicity configurations. Excluding the terms emerging from the $n=1,3$ core factors -- as explained in the text -- we find agreement with the number of crossing-symmetric contact terms at $\cO(\mathfrak{a}^{4})$ for the opposite-helicity amplitude in ref.~\cite{Bautista:2022wjf}. At $\cO(\mathfrak{a}^{6})$, ref.~\cite{Bautista:2022wjf} has some terms with the $n=1,3$ core factor. Our set contains all their contact terms.}
    \label{tab:NumberEvenDissContactTerms}
\end{table}
\begin{table}
    \centering
    \begin{tabular}{c|c|c}
         & general & crossing-symmetric \\
        \hline
        \hline
        $n=4$ & $\frac{1}{6}k(k-1)(2k-1)$ & $\frac{1}{6}k(k^{2}-1)$ \\
        \hline
        $n=3$ & $\frac{1}{3}k(k^{2}-1)$ & $\frac{1}{6}k(k^{2}-1)$ \\
        \hline
        $n=2$ & $(k-1)^{2}$ & $\frac{1}{2}k(k-1)$ \\
        \hline
        $n=1$ & $k(k-1)$ & $\frac{1}{2}k(k-1)$ \\
        \hline
        $n=0$ & $(k-1)^{2}$ & $\frac{1}{2}k(k-1)$ \\
        \hline
        \hline
        Total & $\frac{1}{6}(k-1)(4k^{2}+19k-12)$ & $\frac{1}{6}k(k-1)(2k+11)$
    \end{tabular}
    \caption{The total number of independent $\cO(\mathfrak{a}^{2k+1\geq5})$ dissipative contact term coefficients, and the number left over after requiring crossing symmetry. Valid for all helicity configurations. Excluding the terms emerging from the $n=1,3$ core factors -- as explained in the text -- we find agreement with the number of crossing-symmetric contact terms at $\cO(\mathfrak{a}^{5})$ for the opposite-helicity amplitude in ref.~\cite{Bautista:2022wjf}.}
    \label{tab:NumberOddDissContactTerms}
\end{table}

\section{Summary \& outlook}\label{sec:Summary}

We have shed light on subtleties that must be accounted for when recursively computing classical amplitudes with virtual massive spinning particles.
Specifically, BCFW recursion can produce intermediate expressions which are superclassical at leading order in $\hbar$, thus demanding that one keeps track of subleading-in-$\hbar$ effects in order to completely construct classical amplitudes.
Recursively constructing the classical gravitational Compton amplitude for all helicity configurations has illustrated this necessity.

In the opposite-helicity case, we combined BCFW recursion with the technique presented in ref.~\cite{Aoude:2022trd} for removing unphysical poles to write a local classical amplitude for a general spinning compact object to all orders in its classical spin vector.
When gravitons of the same helicity were scattered, we were able to write an amplitude with no unphysical poles up to fifth order in spin.
Up to third order in spin, both helicity configurations agree with the results of the classical computation of ref.~\cite{Saketh:2022wap}.
At fourth order, our results are in agreement with the amplitude derived from the action of ref.~\cite{Bern:2022kto}.

To complete the Compton amplitude, we also counted and explicitly wrote down all independent contact terms that could potentially contribute to black-hole scattering at 2PM, including both conservative and dissipative effects.
We counted the contact terms both with and without the crossing symmetry imposed in ref.~\cite{Bautista:2022wjf}, and found agreement with their counting in the conservative sector.
In the dissipative sector, our space of contact terms contains, and is larger than, the space of contact terms needed in ref.~\cite{Bautista:2022wjf} to match to solutions of the Teukolsky equation.

Our analysis has uncovered two differences between the Compton amplitude pertaining to Kerr black holes compared with general compact objects.
First, the former can be constructed recursively using only leading-in-$\hbar$ information at all steps in the computation, since all $\cO(\hbar\times 1/\hbar)$ effects cancel within each factorization channel.
Second, and more similar to the minimal coupling condition of ref.~\cite{Arkani-Hamed:2017jhn}, the same-helicity amplitude exhibited the best massless-limit behavior above linear order in spin in the black-hole case.
Above cubic order in spin, the general-object amplitude was divergent as $m\rightarrow0$ at the spins considered, and the black-hole limit was required to quell this divergence.
This makes clear the influence of minimal coupling at three points on higher-point amplitudes, and indicates a potential extension of the notion of minimal coupling to higher multiplicities.
In particular, the form of the three-point amplitude in \cref{eq:ThreePointPositive,eq:ThreePointNegative} hides the significance of the coefficient values $C_{S^{j}}=1$, which is elucidated again by considering the higher-point amplitude.
It is conceivable, then, that considering the three-graviton-emission amplitude will suggest values for the contact-term coefficients in \Cref{sec:Contact} that improve the massless limit of the higher-point amplitude.

Such a method for assigning values to contact-term coefficients is not without its difficulties, however, primarily of which is the likely occurrence of non-localities in higher-point amplitudes at high spin, which must be removed.
Second of all, the non-commutativity of the classical limit with BCFW recursion means the construction of higher-point amplitudes in terms of classical quantities becomes cumbersome, necessitating tracking ever-more subleading parts of lower-point amplitudes.
Finally, any effective (i.e. amplitudes) determination of contact-term coefficients can only be interpreted as describing Kerr black holes insofar as it matches general-relativistic computations, such as those in refs.~\cite{Dolan:2008kf,Levi:2015msa,Saketh:2022wap,Bautista:2022wjf}.

Nevertheless, it is crucial to identify as many differences as possible between black-hole and general-object amplitudes in the pursuit of an amplitudes-based understanding of Kerr black holes.

\acknowledgments

Four-vector manipulations were performed using \texttt{FeynCalc} \cite{MERTIG1991345,Shtabovenko:2016sxi,Shtabovenko:2020gxv}.
I am grateful in particular to Lucile Cangemi, Henrik Johansson, and Andres Luna for very helpful discussions about this work.
I would also like to thank Francesco Alessio, Rafael Aoude, Fabian Bautista, Alessandro Georgoudis, Andreas Helset, Paolo Pichini, and Justin Vines for stimulating discussions about this and related topics.
Furthermore, I thank Andres Luna and Fei Teng, and Muddu Saketh and Justin Vines for sharing unpublished versions of their Compton amplitudes for comparison.
For comments on the manuscripts, I thank Rafael Aoude, Andreas Helset, Henrik Johansson, Jung-Wook Kim, and Andres Luna.
I am grateful to Nordita for their ongoing hospitality.
This work is supported by the Knut and Alice Wallenberg Foundation under grants KAW 2018.0116 (From Scattering Amplitudes to Gravitational Waves) and KAW 2018.0162.

\appendix

\section{Spin vector conventions and properties}\label{app:SpinProperties}

The ring radius $a^{\mu}$ acting on a general spin representation is related to the spin tensor $S^{\mu\nu}$ in that representation by identifying it with the Pauli-Lubanski pseudovector:
\begin{align}
    a^{\mu}&=-\frac{1}{2m}\epsilon^{\mu\nu\alpha\beta}v_{\nu}S_{\alpha\beta}.
\end{align}
This relation can be inverted to express the spin tensor in terms of the ring radius, since $v_{\mu}S^{\mu\nu}=0$:
\begin{align}
    S^{\mu\nu}=-m\epsilon^{\mu\nu\alpha\beta}v_{\alpha}a_{\beta}.
\end{align}
Products of spin vectors in the spin-$s$ representation are related to products in the spin-$1/2$ representation through
\begin{align}
    {\left(a^{\mu_{1}}_{s}\dots a^{\mu_{n}}_{s}\right)_{\alpha_{1}\dots\alpha_{2s}}}^{\beta_{1}\dots\beta_{2s}}&=\frac{(2s)!}{(2s-n)!}{(a^{\mu_{1}}_{1/2})_{\alpha_{1}}}^{\beta_{1}}\dots {(a^{\mu_{n}}_{1/2})_{\alpha_{n}}}^{\beta_{n}}{\delta_{\alpha_{n+1}\dots\alpha_{2s}}}^{\beta_{n+1}\dots\beta_{2s}}+\dots,
\end{align}
where the $+\dots$ represents terms which are subleading in $\hbar$.
Every term at next-to-leading-order in this conversion is antisymmetric in exactly two Lorentz indices.
So, if the tensor contracted into this relation is totally symmetric, the subleading terms are suppressed by one more power of $\hbar$.

It can be much simpler to work in the spin-1/2 representation since only one ring radius vector in this representation can appear between a set of spinors.
In the spin-1/2 representation, the ring radius acts on irreps of $SL(2,\mathbb{C})\times SL(2,\mathbb{C})$ as
\begin{align}
    {(a^{\mu}_{1/2})_{\alpha}}^{\beta}&=\frac{1}{4m}\left[\left(\sigma^{\mu}\right)_{\alpha\dot\alpha}v^{\dot\alpha\beta}-v_{\alpha\dot\alpha}\left(\bar{\sigma}^{\mu}\right)^{\dot\alpha\beta}\right], \\
    {(a^{\mu}_{1/2})^{\dot\alpha}}_{\dot\beta}&=-\frac{1}{4m}\left[\left(\bar{\sigma}^{\mu}\right)^{\dot\alpha\alpha}v_{\alpha\dot\beta}-v^{\dot\alpha\alpha}\left(\sigma^{\mu}\right)_{\alpha\dot\beta}\right],
\end{align}
which we have used to derive \cref{eq:Spin1/2Reduction} and its analog for same-helicity scattering.
These relations can be extended to higher spin representations and powers by first converting the spin vectors in each little group space (i.e. on each side of the cut) to the spin-$1/2$ representation, then projecting the product of spin-$1/2$ spin vectors onto a symmetric product of spin vectors in the appropriate spin representation.
For a general polarization sum we find, for example,
\begin{align}
    &\langle\bbv v_{I}\rangle^{\odot(2s-n)}\langle\bbv|a^{\mu_{1}}_{s}\dots a^{\mu_{n}}_{s}|v_{I}\rangle^{\odot n}[v^{I}\bv]^{\odot(2s-k)}[v^{I}|a^{\nu_{1}}_{s}\dots a^{\nu_{k}}_{s}|\bv]^{\odot k} \\
    &=\langle\bbv|^{2s}\{a^{\mu_{1}}_{s},\dots,a^{\mu_{n}}_{s},a^{\nu_{1}}_{s},\dots,a^{\nu_{k}}_{s}\}|\bv\rangle^{2s}-\frac{ink}{2m^{2}}\langle\bbv|^{2s}\{S^{\mu_{1}\nu_{1}}_{s},a^{\mu_{2}}_{s},\dots a^{\mu_{n}}_{s},a^{\nu_{2}}_{s},\dots,a^{\nu_{k}}_{s}\}|\bv\rangle^{2s},\notag
\end{align}
up to sub-subleading corrections in $\hbar$.
We have assumed that the $\mu_{i}$ are all contracted with one four-vector and the $\nu_{i}$ with another.
For fixed $n,\,k$, the first term always appears for high enough total spin $s$.
The numerator of the second term is determined combinatorially, simply by writing out the little group symmetrizations explicitly.
For $s=1/2$ and $n=k=1$, we recover \cref{eq:Spin1/2Reduction}.

\section{Covariantization of classical results}\label{app:ClassicalComparison}

The authors of ref.~\cite{Saketh:2022wap} computed the amplitude for the scattering of a gravitational plane wave off of a general compact object up to cubic order in the object's spin vector.
They expressed their amplitudes for polar scattering using the four-vectors
\begin{align}
    k^{\mu},&\quad l^{\mu},\notag \\
    w_{S}^{\mu}&=\frac{1}{2\omega\cos^{2}(\theta/2)}\left[\omega(k^{\mu}+l^{\mu})-i\epsilon^{\mu\nu\alpha\beta}k_{\nu}l_{\alpha}v_{\beta}\right], \\
    w_{O}^{\mu}&=-\frac{1}{2\omega\sin^{2}(\theta/2)}\left[\omega(k^{\mu}-l^{\mu})+i\epsilon^{\mu\nu\alpha\beta}k_{\nu}l_{\alpha}v_{\beta}\right],\notag
\end{align}
where the scattering angle of the plane wave is denoted by $\theta$.
To match to $\cM_{++}$ and $\cM_{+-}$ in ref.~\cite{Saketh:2022wap} we must take $k^{\mu}=-q_{3}^{\mu}$ and $l^{\mu}=q_{4}^{\mu}$, which leads to
\begin{subequations}
    \begin{align}
        \frac{t_{14}-t_{13}}{2}\frac{w^{\mu}}{y}&=-\bar{w}_{S}^{\mu}-\frac{(t_{14}-t_{13})s_{34}}{4(t_{13}t_{14}-m^{2}s_{34})}mv^{\mu}+\cO(\hbar^{2}), \\
        \frac{t_{14}-t_{13}}{2}\frac{w_{++}^{\mu}}{y_{++}}&=w^{\mu}_{O}+\frac{t_{14}-t_{13}}{4}\frac{v^{\mu}}{m}+\cO(\hbar^{2}).
    \end{align}
\end{subequations}
The spin-supplementary condition $v\cdot\mathfrak{a}=0$ makes it so that $\bar{w}_{S}^{\mu}$ and $w_{O}^{\mu}$ encode the non-vanishing parts of the contractions of $w^{\mu}$ and $w_{++}^{\mu}$ with $\mathfrak{a}^{\mu}$.
The contractions of the spin with our helicity vectors $w^{\mu}$, $\bar{w}^{\mu}$, and $w_{++}^{\mu}$ are related by
\begin{subequations}\label{eq:wTow++}
    \begin{align}
        \frac{t_{14}-t_{13}}{2}\frac{w\cdot\mathfrak{a}}{y}&=\frac{1}{m^{2}s_{34}-t_{13}t_{14}}\left[t_{13}t_{14}(q_{4}\cdot\mathfrak{a})-m^{2}s_{34}\frac{t_{14}-t_{13}}{2}\frac{w_{++}\cdot\mathfrak{a}}{y_{++}}\right], \\
        \frac{t_{14}-t_{13}}{2}\frac{\bar{w}\cdot\mathfrak{a}}{\bar{y}}&=\frac{1}{m^{2}s_{34}-t_{13}t_{14}}\left[-t_{13}t_{14}(q_{3}\cdot\mathfrak{a})+m^{2}s_{34}\frac{t_{14}-t_{13}}{2}\frac{w_{++}\cdot\mathfrak{a}}{y_{++}}\right],
    \end{align}
\end{subequations}
which were useful in the restoration of locality to the same-helicity amplitude.

For the purposes of comparison with the results in \Cref{sec:Compton}, it is useful to covariantize the amplitudes of ref.~\cite{Saketh:2022wap}.
Accounting for the helicity weights of the amplitude, it is possible to do so uniquely in this case.
Their helicity-preserving amplitude is covariantized by the expansion of \cref{eq:OppositeHelicityCompton} up to third order in spin.
In the helicity-reversing case, their amplitude is covariantized as\footnote{We have had to switch the sign on the exponential in eq.~(5.9) of ref.~\cite{Saketh:2022wap} in order to obtain full agreement with our results above.}
\begin{align}\label{eq:S&VLocal}
    \cM_{+-}&=\frac{y_{++}^{4}}{t_{13}t_{14}s_{34}}\left\{\exp\left[(q_{3}+q_{4})\cdot\mathfrak{a}\right]\frac{}{}\right. \\
    &+\frac{C_{S^{2}}-1}{2!}\left[(q_{3}+q_{4})\cdot\mathfrak{a}]^{2}-\frac{s_{34}}{2}\left[\mathfrak{a}^{2}+4m^{2}\left(\frac{w_{++}\cdot\mathfrak{a}}{y_{++}}\right)^{2}\right]\right]\notag \\
    &+\frac{C_{S^{2}}-1}{2!}\frac{s_{34}}{4}\left[(q_{3}+q_{4})\cdot\mathfrak{a}+3(t_{14}-t_{13})\frac{w_{++}\cdot\mathfrak{a}}{y_{++}}\right]\left[\mathfrak{a}^{2}+4m^{2}\left(\frac{w_{++}\cdot\mathfrak{a}}{y_{++}}\right)^{2}\right]\notag \\
    &+(C_{S^{2}}-1)^{2}\frac{s_{34}}{4}\frac{t_{14}-t_{13}}{2}\frac{w_{++}\cdot\mathfrak{a}}{y_{++}}\left[\mathfrak{a}^{2}+4m^{2}\left(\frac{w_{++}\cdot\mathfrak{a}}{y_{++}}\right)^{2}\right]\notag \\
    &\left.+\frac{C_{S^{3}}-1}{3!}\left[[(q_{3}+q_{4})\cdot\mathfrak{a}]^{3}-\frac{3s_{34}}{4}\left[(q_{3}+q_{4})\cdot\mathfrak{a}+(t_{14}-t_{13})\frac{w_{++}\cdot\mathfrak{a}}{y_{++}}\right]\left[\mathfrak{a}^{2}+4m^{2}\left(\frac{w_{++}\cdot\mathfrak{a}}{y_{++}}\right)^{2}\right]\right]\right\},\notag
\end{align}
where we've used \cref{eq:SameHelicityGram} to write the amplitude in a manifestly local form.

\section{Example of removal of unphysical poles in same-helicity amplitude}\label{app:LocalityExample}

We illustrate the removal of poles in $y\bar{y}=t_{13}t_{14}-m^{2}s_{34}$ from the crossing-symmetric, same-helicity Compton amplitude in \cref{eq:SameHelicityComptonCrossSym} at quadratic order in spin.
The procedure is very similar at higher spins, only with more steps in the iteration.

First, the problematic part of the same-helicity amplitude at quadratic order in spin is
\begin{align}
    &\frac{y_{++}^{4}(C_{S^{2}}-1)}{(m^{2}s_{34}-t_{13}t_{14})^{2}}\left\{\frac{m^{4}s_{34}}{t_{13}t_{14}}\left[-\left(\frac{t_{14}-t_{13}}{2}\frac{w_{++}\cdot\mathfrak{a}}{y_{++}}\right)^{2}+[(q_{3}+q_{4})\cdot\mathfrak{a}]\left(\frac{t_{14}-t_{13}}{2}\frac{w_{++}\cdot\mathfrak{a}}{y_{++}}\right)\right]\right.\notag \\
    &\left.+\frac{m^{4}s_{34}^{2}[(q_{3}\cdot\mathfrak{a})^{2}+(q_{4}\cdot\mathfrak{a})^{2}]+t_{13}^{2}t_{14}^{2}[(q_{3}+q_{4})\cdot\mathfrak{a}]^{2}-m^{2}s_{34}t_{13}t_{14}[3(q_{3}\cdot\mathfrak{a})^{2}+3(q_{4}\cdot\mathfrak{a})^{2}+2(q_{3}\cdot\mathfrak{a})(q_{4}\cdot\mathfrak{a})]}{2s_{34}t_{13}t_{14}}\right\},
\end{align}
where we have already applied \cref{eq:wTow++}.
The first step is to add non-local contact terms which convert the double pole in $m^{2}s_{34}-t_{13}t_{14}$ into a simple pole.
A semi-systematic way of identifying an appropriate contact term is to swap the factors $m^{2}s_{34}\leftrightarrow t_{13}t_{14}$ in the numerator of each term individually such that the result has no poles on physical factorization channels.
Doing so above leads to the boundary term
\begin{align}
    \mathcal{B}^{\prime}|_{\mathfrak{a}^{2}}=-&\frac{m^{2}y_{++}^{4}(C_{S^{2}}-1)}{(m^{2}s_{34}-t_{13}t_{14})^{2}}\left[-\left(\frac{t_{14}-t_{13}}{2}\frac{w_{++}\cdot\mathfrak{a}}{y_{++}}\right)^{2}\right.\notag \\
    &\qquad\left.+[(q_{3}+q_{4})\cdot\mathfrak{a}]\left(\frac{t_{14}-t_{13}}{2}\frac{w_{++}\cdot\mathfrak{a}}{y_{++}}\right)-\frac{(q_{3}\cdot\mathfrak{a})^{2}+(q_{4}\cdot\mathfrak{a})^{2}}{2}\right],
\end{align}
which can be seen to have no poles on physical factorization channels, and can thus be freely added to the amplitude without affecting its residues on physical poles.
The result of adding the two is
\begin{align}
    &\frac{y_{++}^{4}(C_{S^{2}}-1)}{m^{2}s_{34}-t_{13}t_{14}}\left\{\frac{m^{2}}{t_{13}t_{14}}\left[-\left(\frac{t_{14}-t_{13}}{2}\frac{w_{++}\cdot\mathfrak{a}}{y_{++}}\right)^{2}+[(q_{3}+q_{4})\cdot\mathfrak{a}]\left(\frac{t_{14}-t_{13}}{2}\frac{w_{++}\cdot\mathfrak{a}}{y_{++}}\right)\right]\right.\notag \\
    &\left.+\frac{m^{2}s_{34}[(q_{3}\cdot\mathfrak{a})^{2}+(q_{4}\cdot\mathfrak{a})^{2}]-t_{13}t_{14}[(q_{3}+q_{4})\cdot\mathfrak{a}]^{2}}{2s_{34}t_{13}t_{14}}\right\}.
\end{align}
The double pole has thus been alleviated to a simple pole.

A common feature of the analysis at the spins considered is that, once the non-locality has been reduced to a simple pole, there are no longer enough Mandelstam variables in the numerator to identify a suitable boundary contribution in the way described above.
We got around this by employing the Gram determinant in \cref{eq:SameHelicityGram} to eliminate all powers of $(q_{3}\cdot\mathfrak{a})^{i}(q_{4}\cdot\mathfrak{a})^{j}$ in the amplitude, which has generally allowed us to construct a final boundary contribution.
In this simple case, however, we do not need this step.
Employing the Gram determinant to remove instead the term linear in $w_{++}\cdot\mathfrak{a}$ reveals that the remaining unphysical pole is spurious, and lands us on the result in \cref{eq:S&VLocal}.

\bibliographystyle{JHEP}
\bibliography{NSCompton}

\end{document}